\documentclass{aa}

\usepackage{graphicx}
\usepackage{txfonts}
\usepackage[citecolor=blue, linkcolor=blue, urlcolor = black, colorlinks = true]{hyperref}

\usepackage{graphicx}	% Including figure files
\usepackage{amsmath}	% Advanced maths commands
\usepackage{amssymb}	% Extra maths symbols
\usepackage{xspace}

\usepackage{siunitx}
\usepackage{comment}
\usepackage{textcomp}
\usepackage{bm}
\usepackage{lscape}
\usepackage{xcolor}
\newcommand{\gDor}{$\gamma$~Dor\xspace}

\newcommand{\Msun}{\,M$_{\odot}$\xspace}
\newcommand{\dMsun}{-M$_{\odot}$\xspace}

\newcommand{\xc}{$X_{\rm c}$\xspace}
\newcommand{\fov}{$f_{\rm ov}$\xspace}

\newcommand{\teff}{$T_{\rm eff}$\xspace}

\newcommand{\logg}{$\log g$\xspace}

\newcommand{\pin}{$\Pi_0$\xspace}
\newcommand{\edit}[1]{#1}%{{\color{blue} #1}}
\newcommand{\editt}[1]{#1}%{{\color{red} #1}}
\newcommand{\editf}[1]{#1}%{{\color{red} #1}}

\newcommand{\mesa}{\texttt{MESA}\xspace}
\newcommand{\gyre}{\texttt{GYRE}\xspace}
\newcommand{\ctpo}{\texttt{C-3PO}\xspace}

\begin{document}

   \title{Calibrating angular momentum transport in intermediate-mass stars from gravity-mode asteroseismology }

   \author{J.S.G. Mombarg\inst{1,2}
          }
   \institute{IRAP, Universit\'e de Toulouse, CNRS, UPS, CNES, 14 avenue \'Edouard Belin, F-31400 Toulouse, France\\
              \email{jmombarg@irap.omp.eu}
              \and 
              Institute of Astronomy, KU Leuven, Celestijnenlaan 200D, 3001 Leuven, Belgium
        }

   \date{Received 20 January 2023; Accepted 28 June 2023}
\titlerunning{AM transport in intermediate-mass stars}
\authorrunning{Mombarg}
% \abstract{}{}{}{}{} 
% 5 {} token are mandatory
 
  \abstract
   {The physical mechanisms driving the transport of angular momentum in stars are not fully understood, as current models cannot explain the observed stellar rotation profiles across all stages of evolution.   
   }
  {By making use of pulsating F-type dwarfs, this work aims at (i) observationally calibrating the efficiency of angular momentum transport, assuming a constant uniform viscosity, and (ii) testing how well state-of-the-art rotating stellar models with angular momentum (AM) transport by rotationally-induced processes can explain observed rotation profiles. In both cases, the aim is to simultaneously reproduce the measured near-core rotation and core-to-surface rotation ratio. }
   {Asteroseismic modelling is applied to a sample of seven slowly rotating pulsators, to derive (core) masses and ages from their gravity-mode oscillations. This work focuses on the main sequence, using models that start with an initial uniform rotation frequency at the start of core-hydrogen burning that is a free parameter. Two treatments of AM transport are considered: (i) a constant uniform viscosity, and (ii) rotationally-induced processes (including the Spruit-Tayler dynamo). Next, the initial rotation frequency of each star is derived from the observed present-day near-core rotation frequency for both treatments. }
   {Asteroseismic modelling of gravity mode periods reveals all seven slowly rotating stars (one of which is not further modelled) in the sample to be near the end of core-hydrogen burning. To explain the near-core rotation rate at the inferred age, initial rotation frequencies at the zero-age main sequence need to be below 10 percent of the initial critical break-up frequency. The derived initial rotation frequencies are consistent with previous works. }
   {A diffusive approximation of angular momentum transport can in general explain the observed rotation profiles of the six slowly-rotating F-type dwarfs, for average values of the viscosity between $2 \cdot 10^5$ and $5 \cdot 10^7~{\rm cm}^2\,{\rm s}^{-1}$ or when the viscosity is computed from rotationally-induced mechanisms. Yet, for three stars in the sample, the core-to-surface rotation fraction from rotationally-induced mechanisms is predicted to be higher than observed.    }


   \maketitle
%
%-------------------------------------------------------------------
\section{Introduction}
Rotation is a key element in the evolution of stars \citep{maeder2009}, but a complete description of the transport of angular momentum (AM) is still lacking in stellar evolution theory for stars born with a convective core and radiative envelope. The photometric data from space missions with a long time base such as NASA's {\it Kepler} \citep{borucki2010} and TESS CVZ \citep{Ricker2015} missions, allowed for the detection of stellar pulsation frequencies of gravity (g), pressure (p), and inertial modes to a high precision. This resulted in many measurements of (near-) core and/or surface rotation frequencies in pulsating dwarfs \citep[e.g.][]{VanReeth2016, VanReeth2018, christophe2018, Li2019, Li2020, Pedersen2021}, red giants \citep[e.g.][]{Deheuvels2014, gehan2018}, and white dwarfs \citep[e.g.][]{hermes2017}, good for a total of more than 1,200 stars \citep[see][for a review]{Aerts2019-ARAA}. In addition, from observed rotationally split multiplets, a rotation profile can be inferred by means of inversion techniques \edit{when the number of observed modes is large enough} \citep[][]{Deheuvels2012, Triana2015, DiMauro2016}. In the low-mass star regime, a transition in the rotation rates due to the Kraft break \citep{Kraft1967} is observed. Stars with effective temperatures below the Kraft break have thick convective envelopes to support a magnetic dynamo field that drives AM loss. This paper focuses on stars of spectral type F, which are thought to have convective envelopes that are too thin to drive efficient magnetic braking. A complete theory of AM transport and of the rotation profiles at birth could possibly allow for stellar age-dating from measured rotation periods \citep[gyrochronology,][]{Skumanich1972} of stars above the Kraft break, where magnetic braking is not efficient.

State-of-the-art stellar evolution models, including those computed in this work, typically include several rotationally-induced (magneto-)hydrodynamical instabilities inducing turbulence that drives transport of AM. \cite{Eggenberger2005} showed that the flat rotation profile measured in the radiative zone of the Sun could be explained by models with AM transport via the magnetic Tayler instability \citep{Tayler1973} as described by \cite{Spruit2002}. Yet, this Spruit-Tayler dynamo was found to be unable to explain the measured core rotation rates in red giants \citep{cantiello2014}.  In general, the confrontation between core rotation frequencies measured in red giants and those predicted from theory revealed that the current physical descriptions of AM transport predict too high core rotation frequencies \citep[e.g.][]{Eggenberger2012, Marques2013, Tayar2013, cantiello2014}. The revised formalism by \cite{fuller2019} based on the magnetic Tayler instability results into more efficient AM transport, and is able to explain the rigid rotation profiles of main-sequence (MS) stars, as well as the rotation frequencies of low-mass red giants and white dwarfs. In addition, the implementation of magneto-rotational evolution by \cite{Takahashi2021} gives a generally good agreement of observations of Ap/Bp stars and asteroseismic measurements of red giants. Yet, the revised magnetic Tayler instability by \cite{fuller2019}, which has a free parameter, cannot simultaneously explain the measured rotation frequencies in subgiants and red giants. It is argued that this formalism cannot be the complete explanation for the current discrepancies between the observed and predicted rotation rates \citep{Eggenberger2019II, DenHartogh2020}. Furthermore, the predicted timescales of AM transport in main-sequence stars by \cite{fuller2019} seems to be shorter than those inferred from observations of solar-like stars in stellar clusters \citep{Gallet2013}. One of the aims of this work is to test whether rotationally-induced processes alone are sufficient to explain AM transport in main-sequence stars that have a convective core and radiative envelope and are close to the end of core-hydrogen burning. 

Efficient transport of AM can, however, also take place through internal gravity waves (IGWs) excited at the boundary of the convective core (or at the base of convective envelope in low-mass stars) \citep{Charbonnel2005,Rogers2013}. The efficiency is dependent on the generated wave spectrum, and contradictions exist between theoretical works \citep{Goldreich1990} and numerical simulations \citep[e.g.][]{Rogers2013, Edelmann2017}. Additionally, IGWs tend to induce shear, and their effect on the rotation profile cannot be modelled diffusively \citep[see][for a non-diffusive implementation]{Mathis2013}. This work, however, takes a diffusive approach, focusing only on (magneto-)hydrodynamical processes that are induced by rotation, investigating if any additional physics such as IGWs is required in the AM transport of intermediate-mass stars.  

The aforementioned studies that have carried out a comparison between the theory and observations aimed to explain the rotation velocities of a sample of stars as a whole. This work takes a somewhat different approach by testing rotating stellar evolution models on a star-to-star basis, using stars that are well-characterised by asteroseismic modelling. Here, the focus is only on the main-sequence phase of F-type stars, as AM transport already is of importance on the MS \citep[e.g.][]{Ouazzani2019, Pedersen2022b, Moyano2023}, and indications have been found that, for both low-mass and intermediate-mass, strong coupling on the MS is needed \citep{Aerts2019-ARAA}.    

This work makes use of the $\gamma$~Doradus (\gDor) variables \citep{kaye1999}, which have masses between roughly 1.3 and 2.0\Msun. These stars are excellent test beds to confront the state-of-the-art rotating stellar evolution models, as precise near-core rotation frequencies can be measured from the excited g~modes. In addition, the \gDor instability strip partially overlaps with the one of the $\delta$~Scuti variables, and hybrid pulsators have been observed \citep[e.g.][]{Audenaert2022} for which also the envelope rotation frequency can be measured from the p~modes. Moreover, even in the absence of observed p~modes, surface rotation rates of \gDor stars have been measured from rotational modulation, revealing these stars rotate nearly uniformly. 

Most observational studies on AM transport focus on comparing rotation frequencies in different evolutionary stages, or differential rotation in stars past hydrogen burning, while the evolution of the rotational profiles on the main sequence is not well characterised. \cite{Ouazzani2019} focused on a sample of \gDor stars with measured near-core rotation frequencies, and used the buoyancy travel time ($\Pi_0$) as an indicator of age. They conclude that the framework of \cite{Zahn1992} does not reproduce the observations. Yet, \pin is degenerate with respect to mass and age \citep{Mombarg2019}. Therefore, a young star at the lower-mass end of the stability strip, has the same \pin value as an old star at the higher mass end. This implies that when testing AM transport models against a sample of stars with a measured rotation frequency and buoyancy travel time, one has to account for the unknown mass and age, thereby limiting the constraints that can be imposed. Increasing the sample size, particularly with stars that are most likely close to the end of the MS, does improve upon this \citep{Moyano2023}.

The g-mode oscillations follow regular patterns when the difference in the period of modes with the same spherical degree ($\ell$), azimuthal order ($m$) and consecutive radial order ($n$) are plotted as a function of the period itself \citep{Tassoul1980, miglio2008, bouabid2013}. The periods belonging to a period-spacing pattern, where $\Delta P = P_{n+1} - P_{n}$ is plotted against $P_n$, are either found manually \citep[e.g.][]{VanReeth2015a} or in a (semi) automated way \citep[e.g.][]{Garcia2022, Li2020}. From these patterns it is possible to derive the mode identification (i.e. the values for $\ell$ and $m$), such that these periods can be modelled. The work of \cite{Mombarg2021} describes a methodology based on deep learning for inferring the stellar mass, age, and core mass with a relatively high-precision from the identified periods of the observed g~modes, provided enough modes are observed. In this paper, this methodology is applied to a sample of slowly-rotating \gDor stars to study the accuracy of the current rotating one-dimensional stellar structure and evolution models on the main sequence. 

This work aims to test rotating 1D stellar models, by deriving asteroseismic ages for a sample of slowly-rotating \gDor stars with known near-core (and surface) rotation frequencies. We consider two treatments of AM transport: (i) a constant uniform viscosity (Sect.~\ref{sec:nu_nonrot}), and (ii) rotationally-induced processes (Sect.~\ref{sec:nu_rot}). For case (i), the aim is to provide an estimate of the average viscosity needed to reproduce the observations, as to provide a point of calibration for other studies. For case (ii), the is to test whether the viscosity profile predicted from the sum of these rotationally-induced processes can explain the observations. As we only focus on the MS, the physics of the pre-main sequence (PMS) is included by a uniform initial rotation rate at the zero-age main sequence (ZAMS) inferred for each star separately from the present-day near-core rotation frequency (Sect.~\ref{sec:Omegai}). Finally, the conclusions are presented in Sect.~\ref{sec:conclusions}.
  
%--------------------------------------------------------------------
\section{Asteroseismic modelling} \label{sec:modelling}
Keeping in mind the aim of studying the integrated effect of rotation along the main sequence, we focus on slow rotating stars, as these are typically the most evolved. We start from the sample of 22 \gDor stars from \cite{Li2019} that rotate slow enough to observe rotational mode splittings. 

Grid-based asteroseismic modelling becomes quickly computationally expensive when the entire instability strip needs to be sampled with a small enough step size in the fundamental stellar parameters. To remedy this problem, \cite{Mombarg2021} trained a dense neural network on a grid of stellar structure and evolution models, and their pulsation modes to serve as an interpolator. We use their \ctpo\footnote{\url{https://github.com/JMombarg/c3po}} neural network, as well as their methodology to identify radial orders of a series of dipole modes. This method includes the use of the Mahalanobis distance as a merit function to account for the correlations between the observables and theoretical model uncertainties \edit {via the (co)variance matrix of the model grid} \citep{Aerts2018-apjs}. The \ctpo neural network is trained on a grid of non-rotating stellar equilibrium models, where the rotation is taken account at the stage of the stellar pulsation computations. Furthermore, \ctpo is (for now) only trained on prograde dipole modes (i.e. spherical degree $\ell = 1$, and in the convention of positive azimuthal order, $m = 1$), as these are by far the most commonly observed mode geometry \citep{VanReeth2016, Li2020}. Therefore, stars in the sample without observed prograde dipole modes (or too few) are omitted in this work. This concerns 2 out of the 22 stars. Furthermore, we exclude KIC9751996 and KIC11145123, as both have previously been modelled by \cite{mombarg2020}, and KIC11145123 is thought to have undergone binary interaction in the past \citep{Kurtz2014, Takada-Hidai2017}. The methodology described in \cite{Mombarg2021} was applied to a sample of 37 \gDor stars with an effective temperature (\teff), a metallicity ($Z$), and a surface gravity (\logg) derived from spectroscopy \citep{VanReeth2015b}. Yet, for most stars of the sample of \cite{Li2019}, no spectroscopic analysis is available. A cross-match with the sample of \cite{Gebruers2021}, yields one star, KIC7661054, for which \teff, \logg, and $Z$ were derived from spectroscopy. Thus, to reduce the degeneracies present between the fundamental stellar parameters in asteroseismic modelling, the derived luminosity from the Gaia DR2 parallax \citep{Gaia2016} by \cite{Murphy2019} is used. As there are currently no updated distances available for DR3 that are derived in a reliable way \citep{bailer-jones2018}, the distances based of DR2 are used. The asteroseismic modelling is based on evolutionary models for which the luminosity falls within a 3-$\sigma$ confidence interval of the measured Gaia luminosity. 

Starting from the observed periods of the prograde dipole modes, an estimate for the stellar mass ($M_\star$) and hydrogen mass fraction in the core (\xc) is made with the \ctpo neural network. The metallicity is fixed to a solar value of $Z = 0.014$, as spectroscopic studies have shown that the majority of \gDor stars have a metallicity close to that of the Sun \citep[e.g.][]{VanReeth2015b, Kahraman2016, Gebruers2021}. \ctpo assumes a low constant chemical diffusion constant ($D_0$) throughout the radiative envelope fixed to 1\,cm$^2$\,s$^{-1}$. The rotation frequency is uniformly sampled within the uncertainty intervals, using the values measured by \cite{Li2019}. 

This methodology is applied to 17 stars from the \cite{Li2019} sample. The resulting 68-percent-confidence intervals are shown in Table~\ref{tab:theta_NN} in the Appendix. The asteroseismic solution of two stars could not be reconciled with the measured luminosity from Gaia. Therefore, KIC5557072 and KIC5810197 are modelled without any constraints imposed on the luminosity of the best-matching model. In addition, for the binary KIC10080943, with both stars exhibiting g-mode oscillations, the luminosity is not taken into account, but the spectroscopic constraints on the effective temperature and surface gravity for both components from \citet[][their Table~1, using the results for fixed microturbulence]{Schmid2015} are imposed. 

\renewcommand{\arraystretch}{1.5}

\begin{table*}
    \centering
        \caption{Near-core rotation rates, luminosities and mode IDs.}
    \begin{tabular}{llll}
    \hline \hline
    KIC & $f_{\rm rot}\,[{\rm d^{-1}}]$ & $\log L/{\rm L_\odot}$ & Mode IDs ($k,m$)  \\
    \hline
    
4480321 & $0.0070_{-0.0007}^{+0.0007}$ & $1.56 \pm 0.04$ & $(0,-1)$, (0,1), (1,0) \\
5038228 & $0.1594_{-0.0008}^{+0.0008}$ & $1.27 \pm 0.04$ & $(0,-1)$, (0,1), (0,2), (1,0) \\
5810197 & $0.0825_{-0.0006}^{+0.0006}$ & $1.29 \pm 0.07^*$ & $(0,-1)$, (0,1) \\
6937123 & $0.1292_{-0.0004}^{+0.0004}$ & $1.08 \pm 0.03$ & $(0,-2)$, $(0,-1)$, (0,1), (0,2), (1,0), (1,1) \\
9244992 & $0.0148_{-0.0008}^{+0.0009}$ & $1.10 \pm 0.05$ & $(0,-1)$, (0,1), (1,0) \\
10080943(A) & $0.0884_{-0.0008}^{+0.0007}$ & $1.35 \pm 0.04^*$ & $(0,-1)$, (0,1) \\
10080943(B) & $0.1361_{-0.0012}^{+0.0012}$ & $1.35 \pm 0.04^*$ & $(0,-1)$, (0,1), (1,0) \\

\hline
    \end{tabular}
    \tablefoot{Measured near-core rotation rates, luminosities and mode IDs from \cite{Li2019} used for the asteroseismic modelling. Here, the meridional degree $k = \ell - |m|$, as these are g~modes. Luminosities that were not taken into account in the modelling are marked with an asterisk.}
    \label{tab:frot}
\end{table*}

\begin{figure}
    \centering
    \includegraphics[width = \columnwidth]{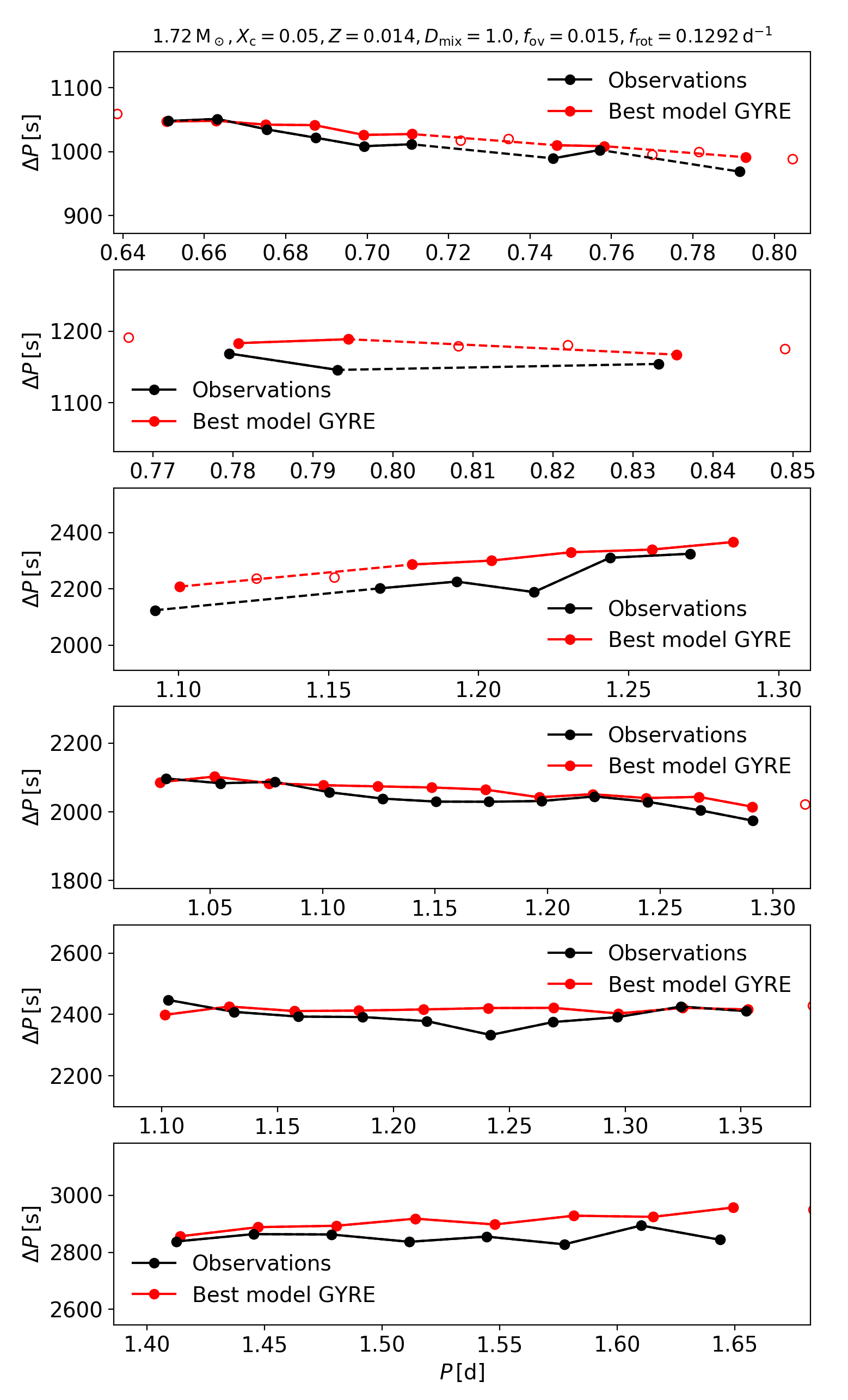}
    \caption{Period-spacing patterns of the best-matching \gyre model for KIC6937123 (in red) and the observed pattern from \cite{Li2019} (in black). The uncertainties on the observations are smaller than the symbol size. Open symbols indicate skipped radial orders in the model. Mode IDs $(\ell, m)$ from top to bottom: (2,2), (2,1), $(2,-2)$, (1,1), (1,0), and $(1,-1)$.}
    \label{fig:PSP69}
\end{figure}

We then select only stars for which a well-constrained initial solution could be achieved from the prograde dipole modes. Table~\ref{tab:frot} lists the input data from \cite{Li2019} for these seven stars. Once the $M_\star$ and \xc have been constrained from the prograde dipole modes with \ctpo, refined stellar evolution models are computed for five evenly sampled masses within the 68-percent-confidence interval, using \mesa r22.05.1 \citep{Paxton2011, Paxton2013, Paxton2015, Paxton2018, Paxton2019, Jermyn2023}. For each mass, tracks are computed with a core-boundary mixing (CBM) parameter \fov (see Eq.~(9) in \citet{Freytag1996}, where the velocity scale height and pressure scale height are related by $h_v = f_{\rm ov}h_P$) varying between 0.005 and 0.035 with a step size of 0.005. Next stellar pulsation modes for all observed mode geometries are computed at five evenly sampled points in \xc within the 68-percent-confidence interval using \gyre v5.2 \citep{Townsend2013, Townsend2018}. Radial orders $-100$ to $-10$ are computed with the (uniform) rotation frequency fixed to the measured value. All observed and identified mode periods are fitted simultaneously to derive a final best-matching model. Table~\ref{tab:theta_sample} lists the parameters of the best-matching \mesa models. The \mesa and \gyre models corresponding to the parameters listed in Table~\ref{tab:theta_sample}, and inlists, are available on Zenodo\footnote{\url{https://doi.org/10.5281/zenodo.7737918}}. Figure~\ref{fig:PSP69} shows the observed period-spacing patterns of KIC6937123 (in black) and the predicted periods of the best-matching model (in red). The results of the other six fits to the period-spacing patterns of the stars can be found Figs.~\ref{fig:PSP44}-\ref{fig:PSP10B} in the Appendix. 

It is known from previous modelling of period-spacing patterns that the mismatch between the predictions of the current pulsation models and the observed mode periods is larger than what is expected from the observational uncertainties \citep[e.g.][]{buysschaert2018, Pedersen2021, Mombarg2021, Szewczuk2022}. This is also the case for the seven stars modelled in this work. The origins of these relatively large \editt{mismatches between theory and observations are} summarised in \cite{Aerts2018-apjs}. 

Interestingly, for the binary KIC10080943, a mass ratio of 0.92 is found, which is close to the value of $0.96 \pm 0.01$ found by \cite{Schmid2015} from fitting the disentangled spectra of both components. Moreover, similar ages ($\tau$) are found while a priori coevality is not enforced, and both components are modelled as single stars. This is justified since the tidal elongation parameter is small, $M_1/M_2(a/R_2)^3 = 0.00027$ (deduced from \cite{Schmid2015}). 

For KIC5038228, KIC9244992, KIC10080943(A), and KIC10080943(B), a constant diffusion coefficient of 1\,cm$^2$~s$^{-1}$ results in predicted period-spacing patterns with dips, which is not observed. These dips are most likely the result of mode trapping in the presence of a chemical gradient located just outside the core boundary \citep{miglio2008}. For these stars, the modelling after the estimate from \ctpo is done again with a higher value for $D_0$, namely 10\,cm$^2$\,s$^{-1}$. Increasing $D_0$ by a factor 10 is sufficient to remove the dips in the period-spacing pattern, while it has a negligible effect on the evolution. The fact that the morphology of the observed period-spacing patterns is not completely reproduced by the models results in a larger degeneracy between mass and age in some cases, as the value of the merit function is mostly determined by the average mode density set by $\Pi_0$ and range of the periods. The inclusion of the inferred luminosity and effective temperature, and/or multiple mode geometries, mitigate this degeneracy.   

\renewcommand{\arraystretch}{1.0}

%\begin{landscape}
\begin{table*}[]
    \centering
        \caption{Parameters of the best-fitting \mesa/\gyre models.  }
    \begin{tabular}{lllllllllllll}
    \hline \hline
    KIC & $M_\star\,[{\rm M}_\odot]$ & $X_{\rm c}$  & $f_{\rm ov}$ & $\log\,D_{0}$ & $\log L/{\rm L_\odot}$ & $\log T_{\rm eff}$ &  $\tau\,[{\rm Gyr}]$ & $m_{\rm cc}/M_\star$ & $R_\star\,[{\rm R}_\odot]$ & $(\Omega/\Omega_{\rm c})_{\rm i}$  \\
    \hline
4480321 & 2.00  & 0.06 & 0.035 & 0 & 1.503 & 3.795 & 1.39 & 0.083 & 4.84 & $\sim$0.01\\
5038228 & 1.83  & 0.08 & 0.035 & 1 & 1.332 & 3.790 & 1.77 & 0.084 & 4.07 & $\sim$0.10\\
5810197 & 1.30  & 0.05 & 0.030 & 0 & 0.681 & 3.762 & 4.85 & 0.067 & 2.19 & \editf{$\sim$0.01}\\
6937123 & 1.72  & 0.05 & 0.015 & 0 & 1.123 & 3.806 & 1.74 & 0.071 & 2.97 & $\sim$0.05\\
9244992 & 1.63  & 0.01 & 0.035 & 1 & 1.142 & 3.772 & 2.57 & 0.066 & 3.55 & $\sim$0.01\\
10080943(A) & 1.95  & 0.10 & 0.015 & 1 & 1.350 & 3.841 & 1.19 & 0.084 & 3.28 &$\sim$0.05\\
10080943(B) & 1.80  & 0.20 & 0.005 & 1 & 1.158 & 3.858 & 1.19 & 0.093 & 2.43 & -\\
\hline
    \end{tabular}

    \label{tab:theta_sample}
\end{table*}
%\end{landscape}
\renewcommand{\arraystretch}{1.5}

\section{AM transport} \label{sec:AM_trans}
The aim of this paper is to calibrate the average efficiency of angular momentum transport and test the prescriptions using \mesa. The stellar structure and evolution models presented in this section take the evolution of the rotation into account, contrary to the non-rotating models used for asteroseismic modelling described in Sect.~\ref{sec:modelling}. Since the stars modelled in this work are slow rotators, it is reasonable to assume the derived stellar parameters are not much affected by neglecting rotation in the stellar equilibrium models and only taking rotation into account for the computation of the mode periods. In \mesa, the transport of angular momentum is treated with a diffusion approximation (Equation (B4) in \cite{Paxton2013}),
\begin{equation}
\begin{split}
    \left(\frac{{\partial \Omega}}{\partial t}\right)_m &= \frac{1}{i}\left( \frac{\partial }{\partial m} \right)_t \left[ (4 \pi r^2 \rho)^2 i \nu_{\rm AM} \left( \frac{\partial \Omega}{\partial m} \right)_t \right] \\
    &- \frac{ \Omega}{r} \left( \frac{\partial r }{\partial t} \right)_m \left(\frac{{\rm d} \ln i }{{\rm d} \ln r} \right),
    \end{split}
    \label{eq:AM_trans}
\end{equation}
where $i$ is the specific moment of inertia of a shell at mass coordinate $m$, and $\rho$ the local density. In this paper, the term `rotational mixing' is used to describe turbulent mixing induced by (magneto-)hydrodynamical processes, and the term `non-rotational' relates to any (unknown) processes modelled by a constant viscosity. The diffusion coefficient (or viscosity) for the transport of angular momentum ($\nu_{\rm AM}$) is the sum of the diffusion coefficient resulting from non-rotational mixing, and the diffusion coefficient from rotation,
\begin{equation}
 \nu_{\rm AM} = f_{\rm \nu, non-rot}\nu_{\rm AM,  non-rot} + f_{\rm \nu, rot}\nu_{\rm AM, rot},
\end{equation}
where $f_{\rm \nu, non-rot}$ and $f_{\rm \nu, rot}$ are factors to scale the different contributions. \editt{In this paper, these two different contributions to the total viscosity are studied separately.} The following processes are included in the computation of $\nu_{\rm AM, rot}$:

\begin{itemize}
    \item Dynamical Shear Instability (DSI): An instability operating on a dynamical timescale, and dependent on the horizontal variation of the velocity profile.
    \item Secular Shear Instability (SSI): When heat loss is taken into account in the criterion of DSI, and operates on a thermal timescale.
    \item Eddington-Sweet (ES) circulation: Large-scale circulations resulting from the fact that in a rotating star the isobars and isotherms do not coincide. \edit{While ES circulation is an advective process, it can be modelled diffusively when it will leads to solid body rotation \citep{Endal1978, Chaboyer1992}. A comparison between the advective and diffusive approach (in massive stars) can be found in the work by \cite{Potter2012}.}
    \item Solberg-H\o iland Instability (SH): An instability linked to the Coriolis force, also taking the stratification into account.
    \item Goldreich-Schubert-Fricke Instability (GSF): A baroclinic instability describing semiconvection and thermohaline mixing \citep{Goldreach1967, Fricke1968, Barker2019}.
    \item Spruit-Tayler dynamo (ST): AM transport via magnetic torques \citep{Spruit2002}. \edit{For given saturated field strengths of the radial and azimuthal components of the magnetic field, an effective viscosity can be derived.}
\end{itemize}
The reader is referred to the works of \cite{Endal1978}, \cite{Pinsonneault1989} and \cite{Heger2000} for more details on these processes. \newline

The diffusion coefficient from rotation is in turn defined as 
\begin{equation}
\begin{split}
    \nu_{\rm AM, rot} &= 
    f_{\rm DSI}  D_{\rm DSI} +
    f_{\rm SSI}  D_{\rm SSI} +
    f_{\rm ES}   D_{\rm ES} +
    f_{\rm SH}   D_{\rm SH} \\
    &+
    f_{\rm GSF}  D_{\rm GSF} +
    f_{\rm ST}   \nu_{\rm ST},
\end{split}
\label{eq:nu_rot}
\end{equation}
where the diffusion coefficients $D$ are computed following the formalism in \cite{Heger2000}, and $f$ are factors to scale the relative contributions.  
\edit{Since the rotational evolution on the PMS is uncertain \citep[e.g.][]{Amard2019}, we focus here only on AM transport on the MS and assume stars reach the ZAMS as solid body rotators.} A precomputed PMS model is in 20 steps relaxed to a model rotating at a certain fraction of the critical \editt{Keplerian} break-up frequency, defined in \mesa as,
\begin{equation}
 \Omega_{\rm c} = \sqrt{\frac{\Gamma G M_\star}{R_\star^3}}.
\end{equation}
Here, $\Gamma = 1 - L_\star/L_{\rm Edd}$, with $L_{\rm Edd}$ the Eddington luminosity. For the stars considered here, $\Gamma \approx 1$. 

In the following sections, we first start by estimating the order of magnitude of the viscosity in the simplest case of a uniform viscosity constant in time to get a sense of the average viscosity that each star has in its present phase. Then, we test whether a viscosity computed from the processes listed above is sufficient to reproduce the observations. Other studies that have tested the theory of AM transport \citep[e.g.][]{Eggenberger2012, Ouazzani2019} combine treatments (i) and (ii), where a uniform viscosity is added to the one predicted from rotationally-induced processes only if these physical processes alone are insufficient to explain the data. However, this work considers the uniform viscosity (i) separately, as this result is easy to translate between different stellar evolution codes and easy to compare with theoretical predictions. 

\subsection{Non-rotational AM transport} \label{sec:nu_nonrot}
As a first step, we set $f_{\rm \nu, non-rot}$ to 1, $f_{\rm \nu, rot}$ to 0, and impose a uniform time-independent value for $\nu_{\rm AM,  non-rot}$. 
The value of $\nu_{\rm AM,  non-rot}$ essentially controls the amount of (radial) differential rotation, where large values result in uniform rotation. By exploiting both the observed pressure modes and gravity modes in both components of the binary KIC10080943, \cite{Schmid2016} were able to infer the rotation frequency in the near-core region and at the surface. For KIC10080943(A), these authors found a core-to-surface rotation ratio between 1 and 1.09, and for KIC10080943(B) a core-to-surface rotation ratio of $0.74 \pm 0.01$. The latter is a peculiar rotation profile, as a core that is rotating slower than the envelope \edit{cannot be realised using Eq.~(\ref{eq:AM_trans}) and starting from uniform rotation. Therefore it is likely that other mechanisms are at play here (e.g. accretion, tidal interactions). } As can be seen from Eq.~(\ref{eq:AM_trans}), diffusive AM transport depends on the radial derivative of $\Omega$, and thus once uniform rotation is reached, the rotation frequency remains constant (or rather only changes due to the change in the inertia profile during the evolution). Therefore, KIC10080943(B) is not modelled further here. Similarly, \cite{Saio2015} derived from the p~modes and g~modes in KIC9244992 a core-to-surface rotation ratio of 1.03. Similar values were also found by \cite{VanReeth2018}, who measured the core-to-surface ratio of eight \gDor stars, and found ratios between 0.95 and 1.05. For five out of the eight stars \cite{Mombarg2021} managed to provide relatively precise age estimates, which tells that these stars have between roughly 25 and 75 percent of their initial hydrogen mass fraction left in the core. While the maximum differential rotation of 1.09 measured in KIC10080943(A) and the study of \cite{VanReeth2018} is based on the near-core rotation rate, \cite{Saio2021} measured the actual core rotation rates derived from the coupling between pure inertial modes and gravito-inertial modes \citep{ouazzani2020}. In their sample of 16 stars, 15 have core rotation frequency between 1 and 1.1 times the rotation frequency of the g-mode cavity. In this paper, these typical limits of radial differential rotation are used as constraints. \newline

\begin{figure}
    \centering
    \includegraphics[width = \columnwidth]{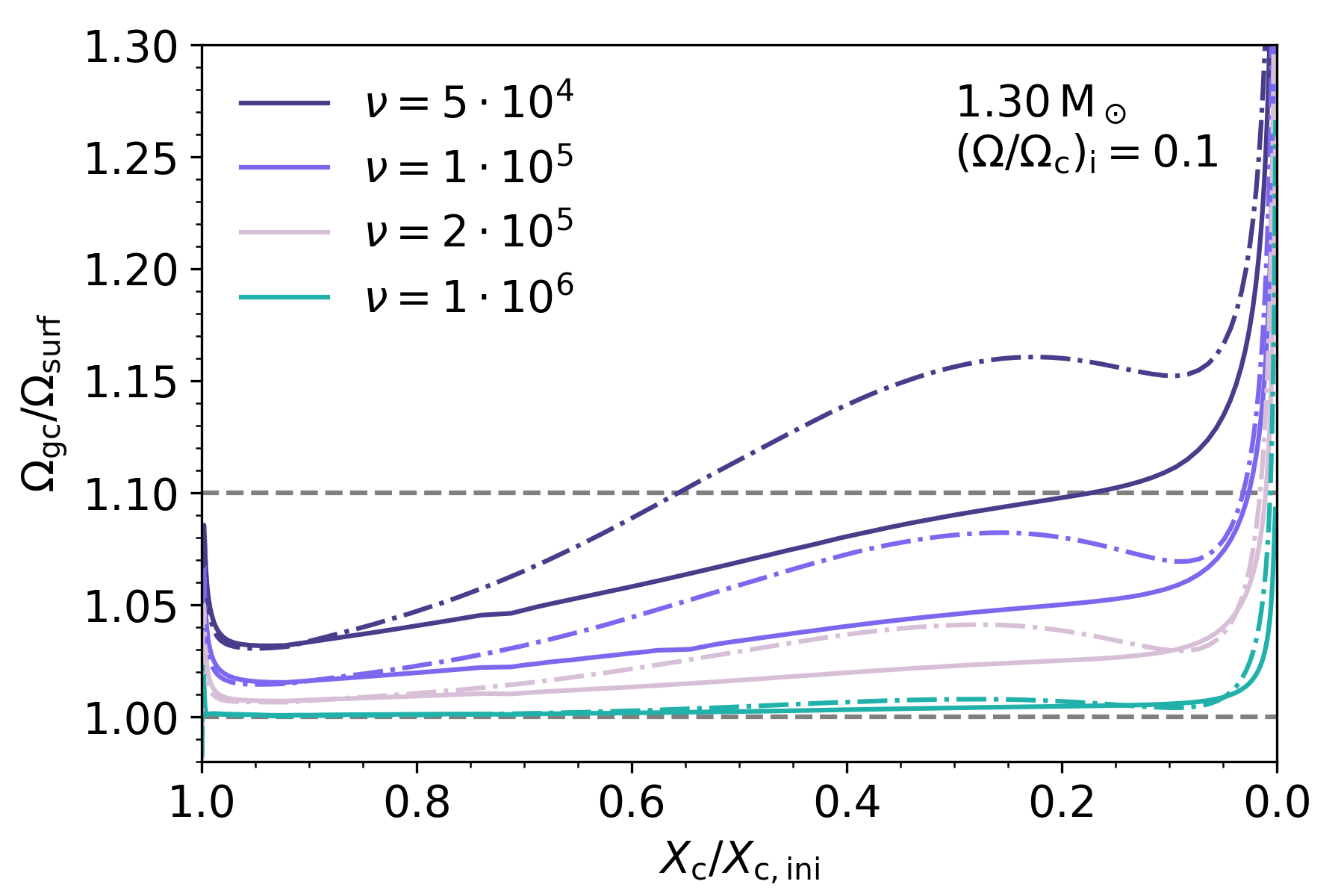}
    \caption{The predicted ratio of the g-mode cavity rotation frequency over the surface rotation frequency accumulated across evolution for a 1.3\dMsun model starting with $(\Omega/\Omega_{\rm c})_{\rm i} = 0.1$. In these models, AM transport is modelled by a constant uniform viscosity, shown in the legend (in cm$^{2}$\,s$^{-1}$). The solid lines are for $f_{\rm ov} = 0.005$, the dashed-dotted lines for $f_{\rm ov} = 0.035$. The typical upper limit of the observed range of core-to-surface rotation is shown by the upper horizontal grey dashed line. The lower grey dashed horizontal line indicates solid body rotation. }
    \label{fig:nu_nonrot_M130_O10}
\end{figure}

\begin{figure}
    \centering
    \includegraphics[width = \columnwidth]{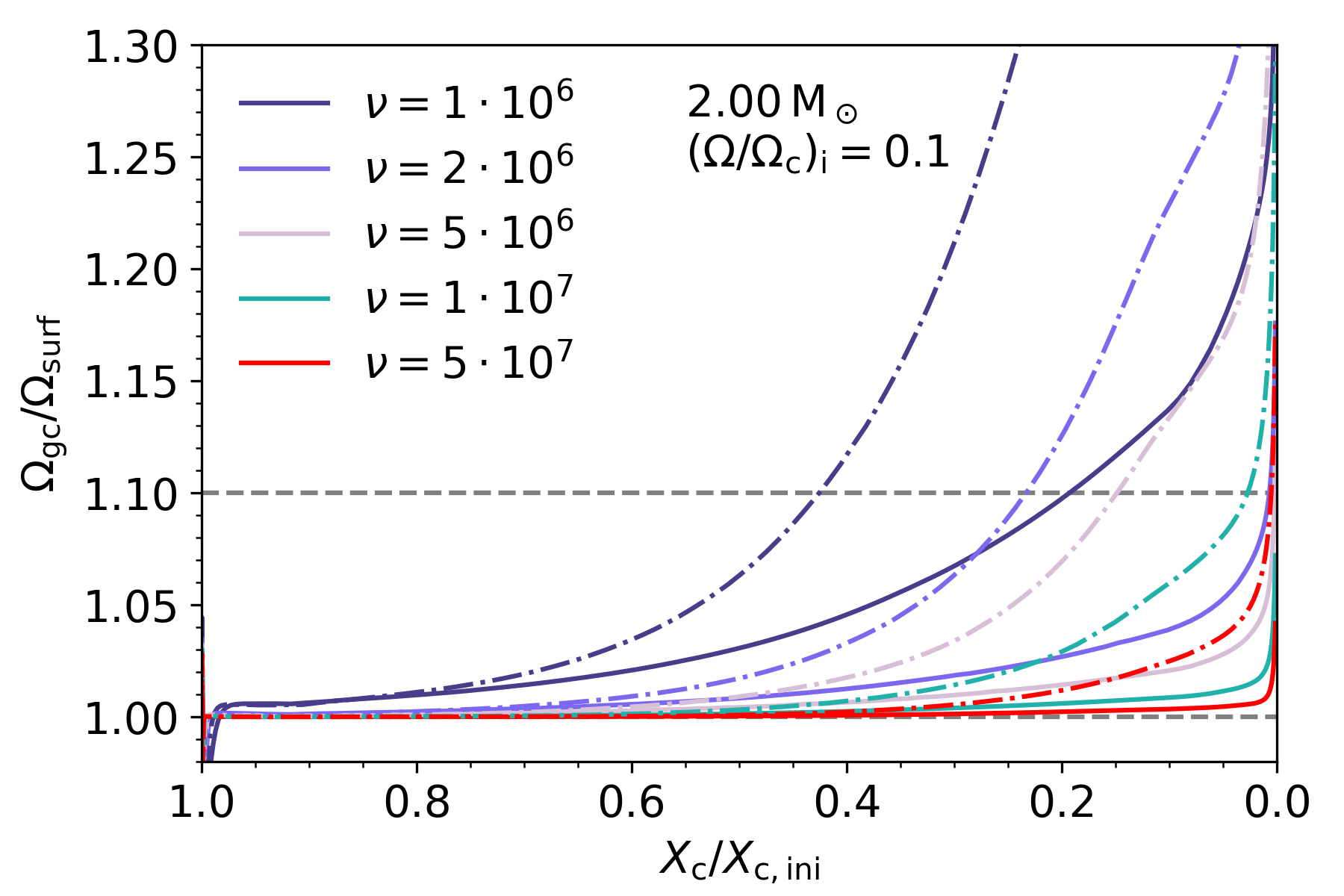}
    \caption{Same as Fig.~\ref{fig:nu_nonrot_M130_O10}, but for a 2\dMsun model. }
    \label{fig:nu_nonrot_M200_O10}
\end{figure}

To make a useful comparison between the near-core rotation rate measured from g~modes and the one in the model, we define the g-mode cavity rotation frequency \citep{Ouazzani2019, Pedersen2022b},
\begin{equation}
    \Omega_{\rm gc} = \frac{\int \Omega(r)N(r)r^{-1}{\rm d}r}{\int N(r)r^{-1}{\rm d}r},
\end{equation}
where $N$ is the Brunt-V\"ais\"al\"a frequency. In the definition of $\Omega_{\rm gc}$, the integrals are evaluated over the region where $N^2 > 0$. Figs.~\ref{fig:nu_nonrot_M130_O10}-\ref{fig:nu_nonrot_M200_O10} show the evolution of $\Omega_{\rm gc}/\Omega_{\rm surf}$ for a 1.3\dMsun and a 2.0\dMsun star, both for \fov equal to 0.005 and 0.035. To ensure that $\Omega_{\rm gc}/\Omega_{\rm surf}$ does not exceed 1.1 on the main-sequence, $\nu_{\rm AM,  non-rot}$ must be higher than $2\cdot 10^5~{\rm cm}^2\,{\rm s}^{-1}$, independent of \fov in case of the 1.3\dMsun model. For the 2.0\dMsun model, $\nu_{\rm AM,  non-rot}$ should be higher than $\sim 2\cdot 10^6~{\rm cm}^2\,{\rm s}^{-1}$ for $f_{\rm ov} = 0.005$ and higher than $\sim 5\cdot 10^7~{\rm cm}^2\,{\rm s}^{-1}$ for $f_{\rm ov} = 0.035$. As is shown in Figs.~\ref{fig:nu_nonrot_M130_O40} and \ref{fig:nu_nonrot_M200_O40} in the Appendix, the evolution of $\Omega_{\rm gc}/\Omega_{\rm surf}$ is only slightly different between the models with an initial rotation rate $(\Omega/\Omega_{\rm c})_{\rm i} = 0.1$ and 0.4. In the 1.3\dMsun models with a larger CBM region (i.e. larger \fov, dashed lines in Fig.~\ref{fig:nu_nonrot_M130_O10}), a local maximum in the ratio $\Omega_{\rm gc}/\Omega_{\rm surf}$ is observed that is related to the slow-down in the expansion of the envelope, which happens more gradual for the model with a smaller CBM region. The envelope of the 2.0\dMsun model expands more rapidly, and thus, the core-to-surface rotation ratio keeps increasing along the MS.  \newline

\begin{figure}
    \centering
    \includegraphics[width = \columnwidth]{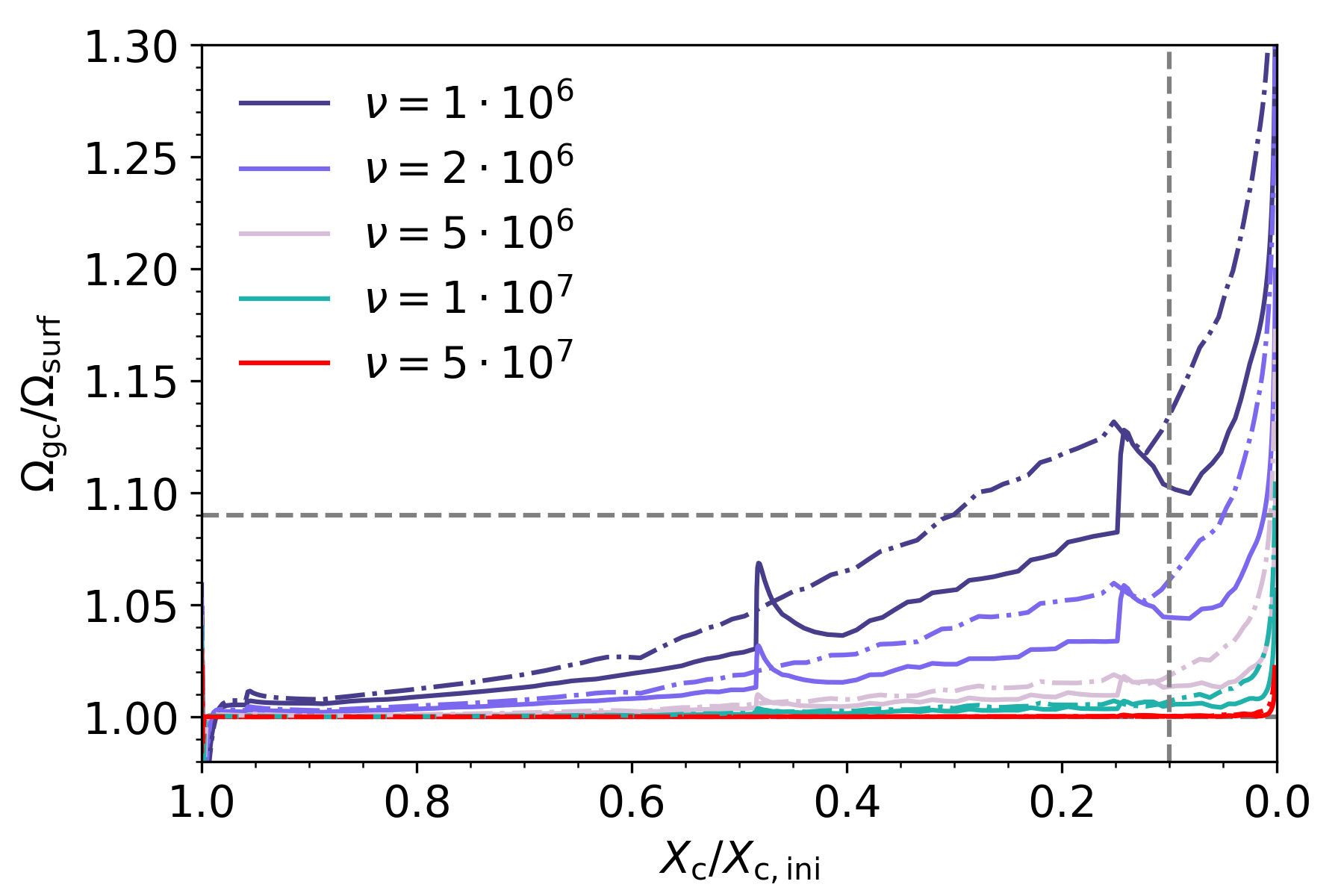}
    \caption{The predicted ratio of the g-mode cavity rotation frequency over the surface rotation frequency as a function of evolution for the best-matching model of KIC10080943(A) listed in Table~\ref{tab:theta_sample}. In these models, AM transport is modelled by a constant uniform viscosity, shown in the legend (in cm$^{2}$\,s$^{-1}$). The solid lines are for $(\Omega/\Omega_{\rm c})_{\rm i} = 0.1$, the dashed-dotted lines for $(\Omega/\Omega_{\rm c})_{\rm i} = 0.4$. \editf{The observed core-to-surface rotation of 1.09 by \cite{Schmid2016} is shown by a grey dashed line.} The vertical grey dashed line marks the age of the best-matching model.
    }
    \label{fig:nu_nonrot_KIC10A}
\end{figure}

\begin{figure}
    \centering
    \includegraphics[width = \columnwidth]{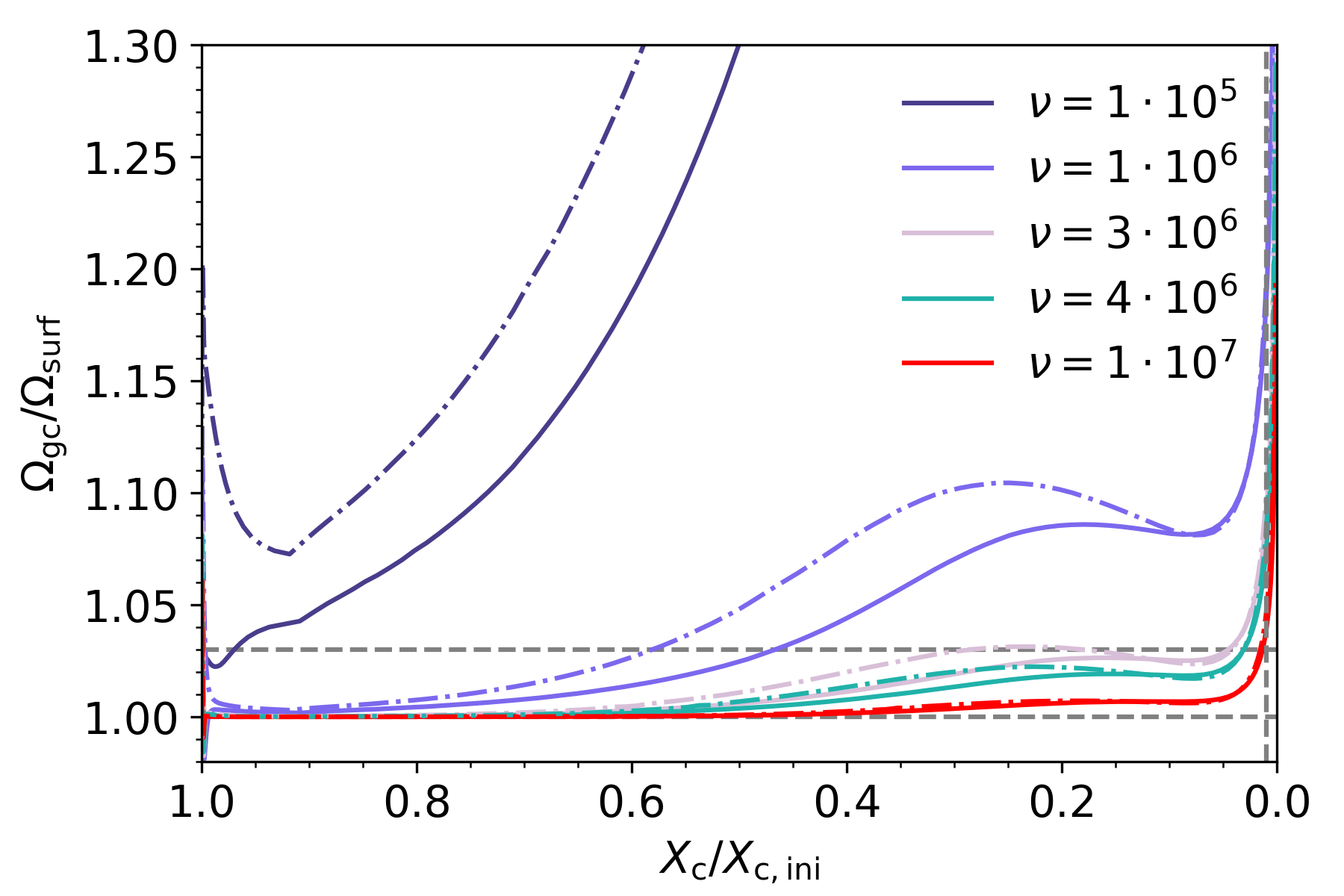}
    \caption{Same as Fig~\ref{fig:nu_nonrot_KIC10A}, but for KIC9244992. \editf{The observed core-to-surface rotation of 1.03 by \cite{Saio2015} is shown by a grey dashed line.} The vertical grey dashed line marks the age of the best-matching model (close to $X_{\rm c}/X_{\rm ini} = 0$).}
    \label{fig:nu_nonrot_KIC92}
\end{figure}

In Fig.~\ref{fig:nu_nonrot_KIC10A}, the evolution of the ratio $\Omega_{\rm gc}/\Omega_{\rm surf}$ is shown for different values of $\nu_{\rm AM,  non-rot}$, where the mass and \fov are fixed to the values of the best-matching model for KIC10080943(A). Assuming a constant viscosity for AM transport, a lower limit of $\nu_{\rm AM,  non-rot} \sim 10^6~{\rm cm}^2\,{\rm s}^{-1}$ is found for KIC10080943(A). As the lower limit of $\Omega_{\rm gc}/\Omega_{\rm surf}$ for this star is consistent with uniform rotation, no upper limit on $\nu_{\rm AM,  non-rot}$ can be inferred.
Since mass and age measurements for KIC9244992 are now available, we also can estimate the value of a constant $\nu_{\rm AM,  non-rot}$ needed to produce a 1.63-${\rm M_\odot}$ star with a core that is rotating about 1.03 times faster than the envelope at $X_{\rm c} \approx 0.01$. The mass of the best-model obtained in this work for KIC9244992 is somewhat higher than 1.45\Msun found by \cite{Saio2015} from modelling the p~modes. However, their best model is consistent with the asteroseismic solutions found by the modelling of the g~modes in this work, but not with the constrains on the measured luminosity (which was not taken into account in the modelling by \citealt{Saio2015}). For the best model in Table~\ref{tab:theta_sample}, we find $\nu_{\rm AM,  non-rot}$ should be of the order of $3 \cdot 10^{6}~{\rm cm}^2\,{\rm s}^{-1}$ to reproduce a core-to-envelope rotation of 1.03 around the estimated age, as is illustrated in Fig.~\ref{fig:nu_nonrot_KIC92}. In summary, when modelling the accumulated effect AM transport in F-type stars with a uniform viscosity, a value between $2\cdot 10^5~{\rm cm}^2\,{\rm s}^{-1}$ and $5\cdot 10^7~{\rm cm}^2\,{\rm s}^{-1}$ is found, where the exact value depends on the stellar mass and CBM efficiency. These ranges are in line with the values of $10^6~{\rm cm}^2\,{\rm s}^{-1}$ found by \cite{Ouazzani2019} for \gDor stars. This is two orders of magnitude larger than the viscosity inferred by \cite{Eggenberger2012} in red giants.  

\subsection{Rotational AM transport} \label{sec:nu_rot}
\begin{figure}
    \centering
    \includegraphics[width = \columnwidth]{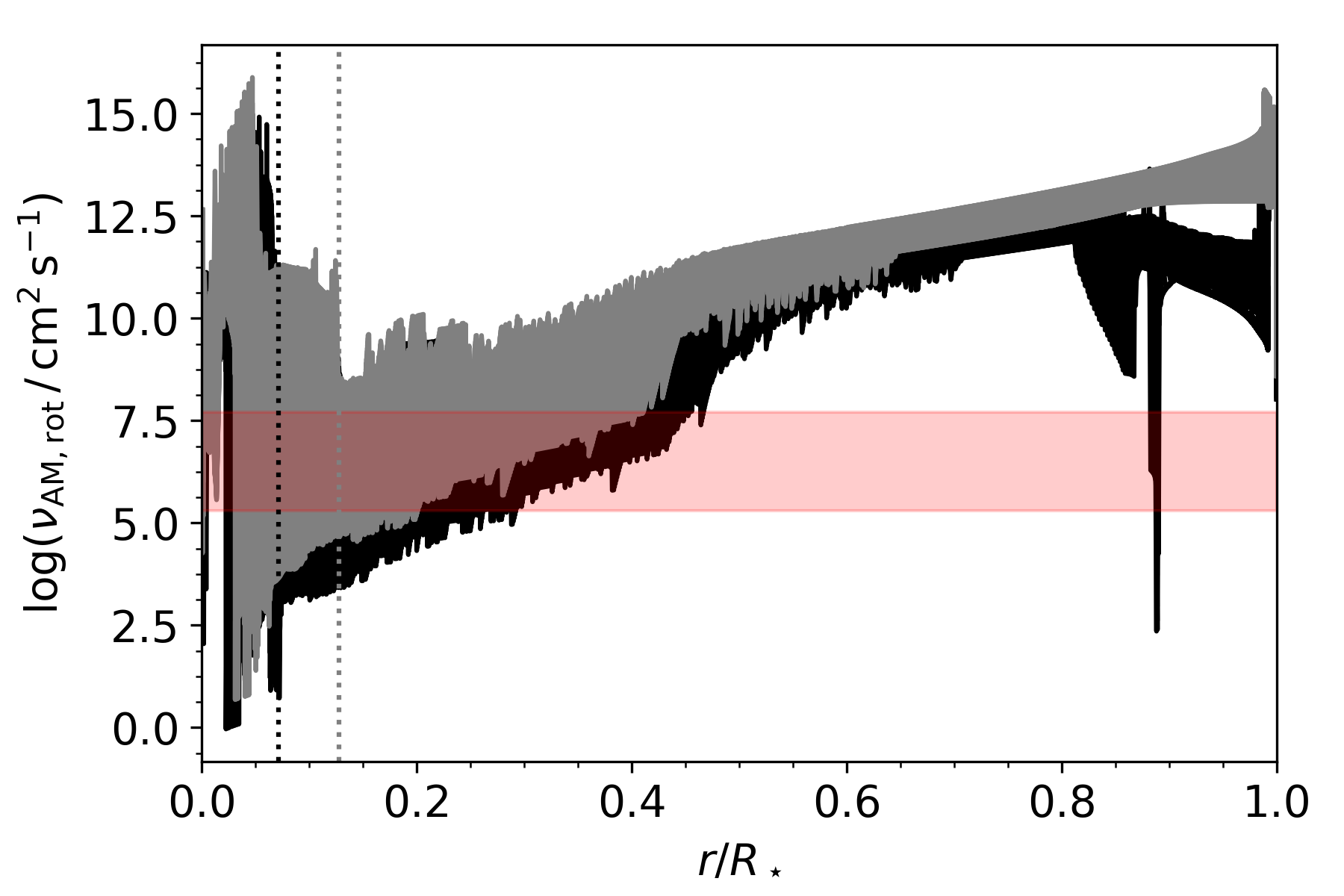}
    \caption{\editt{Range of values of the local viscosity predicted by the rotationally-induced processes discussed in Sect.~\ref{sec:AM_trans} across the main sequence, for 1.3\Msun (in black) and 2\Msun (in grey). For both models $(\Omega/\Omega_{\rm c})_{\rm i} = 0.1$. The red band indicates the range required for a uniform viscosity to match the observations (Sect.~\ref{sec:nu_nonrot}). \editf{The maximum extent of the convective core throughout the evolution is indicated by the dotted lines for both masses.} }}
    \label{fig:nu_rot_prof}
\end{figure}

\begin{figure}
    \centering
    \includegraphics[width = \columnwidth]{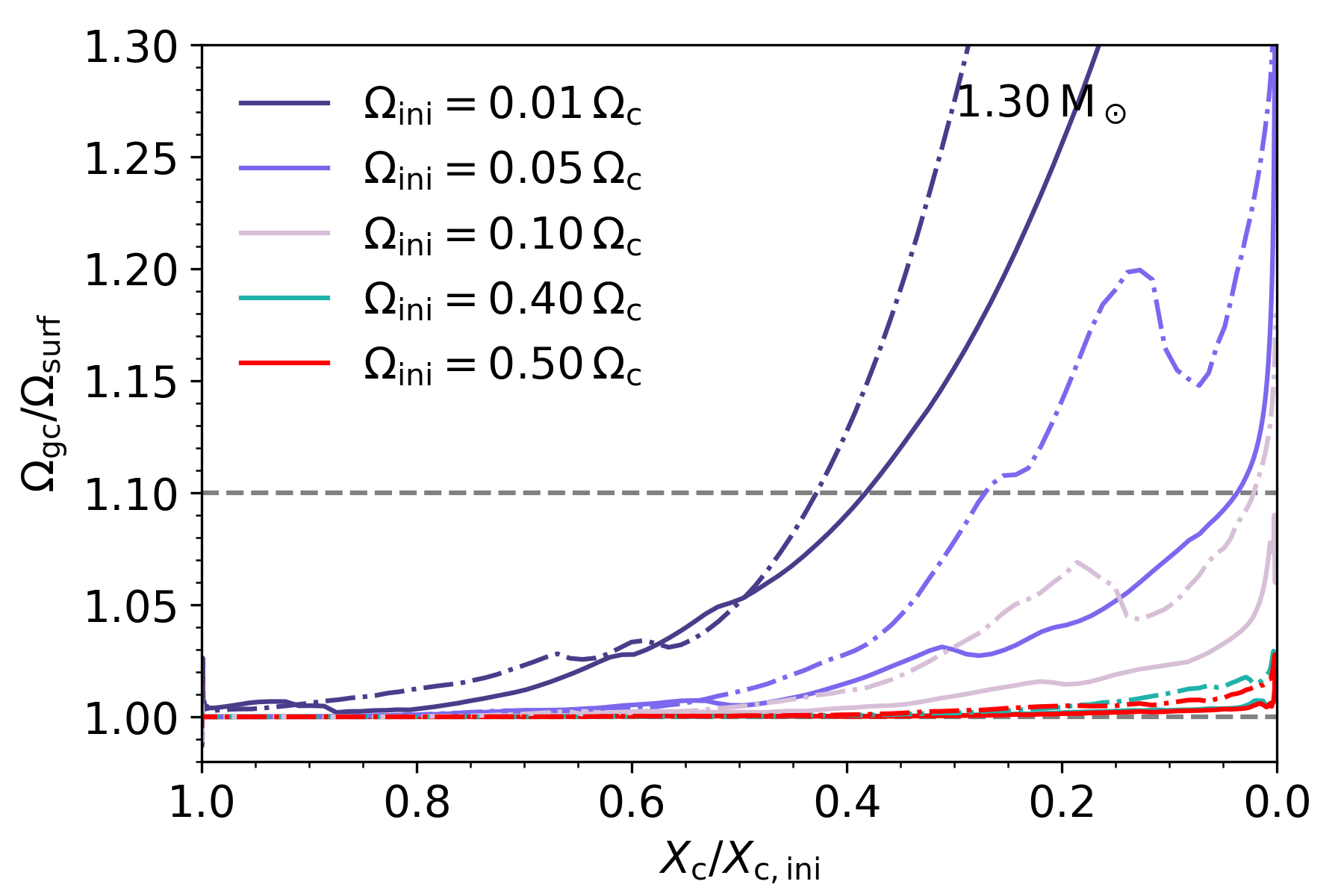}
    \caption{The predicted ratio of the g-mode cavity rotation frequency over the surface rotation frequency as a function of evolution for a 1.3\dMsun model. In these models, AM transport is computed from the rotationally-induced processes described in Eq~(\ref{eq:nu_rot}). The solid lines are for $f_{\rm ov} = 0.005$, the dashed-dotted lines for $f_{\rm ov} = 0.035$. The typical upper limit of the observed range of core-to-surface rotation is shown by the upper horizontal grey dashed line. The lower grey dashed line indicates solid body rotation. }
    \label{fig:nu_rot_M130}
\end{figure}

\begin{figure}
    \centering
    \includegraphics[width = \columnwidth]{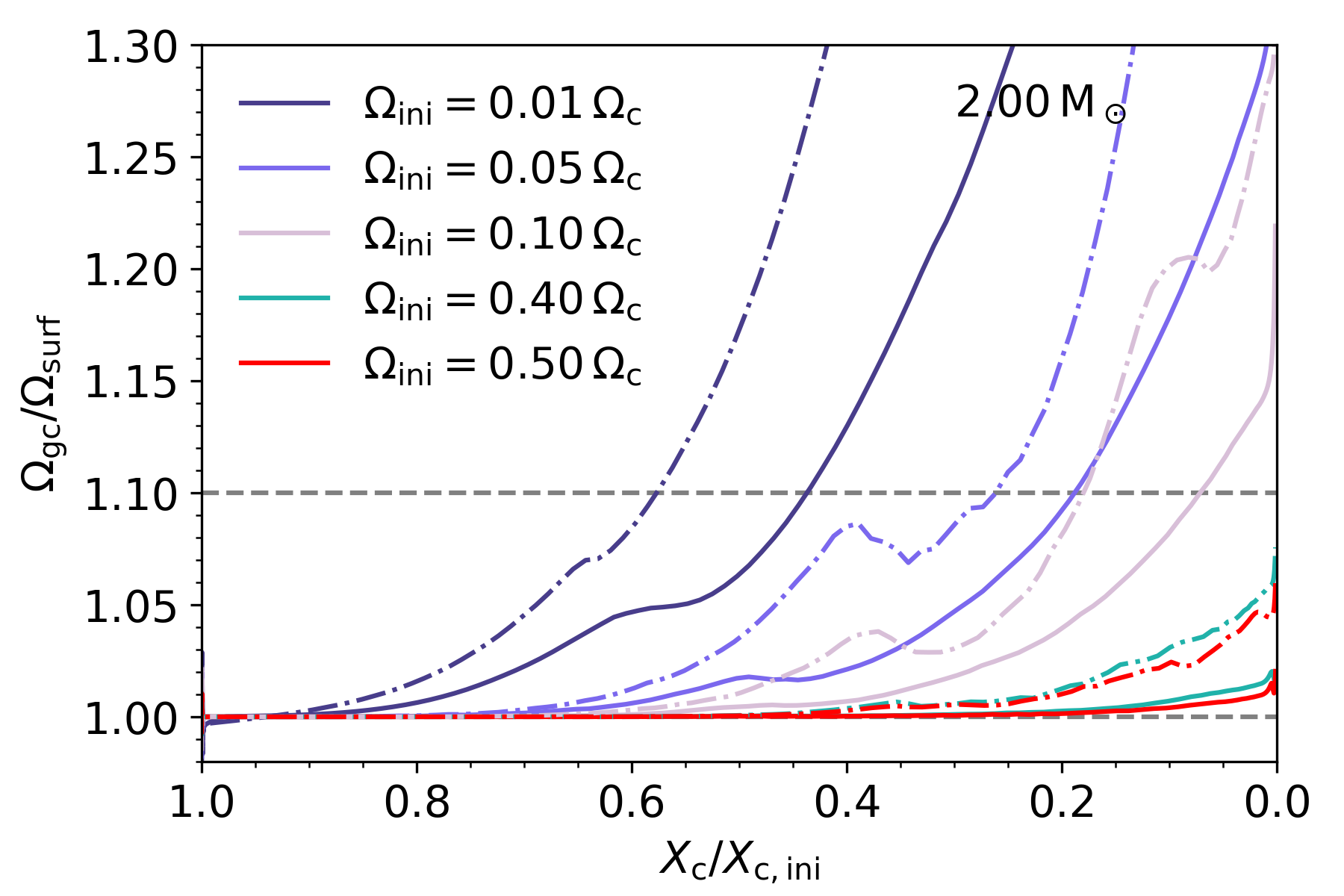}
    \caption{Same as Fig.~\ref{fig:nu_rot_M130}, but for a 2\dMsun model. }
    \label{fig:nu_rot_M200}
\end{figure}

As a next step, we now set $f_{\rm \nu, non-rot}$ to 0, and $f_{\rm \nu, rot}$ to 1. All factors $f$ in Eq.~(\ref{eq:nu_rot}) are set to 1, that is, no ad-hoc scaling of relative contributions of the different processes. In Fig.~\ref{fig:nu_rot_prof}, the range of $\nu_{\rm AM, rot}$ throughout the star is shown for 1.3 and 2.0\Msun along the MS. The transport of AM in the envelope is mostly dominated by Eddington-Sweet circulation and the Spruit-Tayler dynamo. The average values of $\nu_{\rm AM, rot}$ around the core boundary are of the same order of magnitude as the uniform viscosities $\nu_{\rm AM, non-rot}$ inferred in Sect.~\ref{sec:nu_nonrot}. In this case, there is no viscosity parameter to tune. However, the evolution of the $\Omega_{\rm gc}/\Omega_{\rm surf}$ is strongly dependent on the initial rotation frequency, where higher rotation frequencies lead to more efficient AM transport, and thus a lower level of radial differential rotation. As can be seen in Figs.~\ref{fig:nu_rot_M130} and \ref{fig:nu_rot_M200} that show this ratio for 1.3 and 2.0\dMsun respectively, for most models $\Omega_{\rm gc}/\Omega_{\rm surf}$ remains within the range of what is observed \citep[e.g.][]{VanReeth2018, Li2020, Saio2021}. However, for some models with initial rotation frequencies lower than about five percent of the critical frequency, this ratio becomes larger than 1.2 near the end of the main sequence, and is in contrast with what has been observed so far. The sharp variations seen in some models, like the one with $(\Omega/\Omega_{\rm c})_{\rm i} = 0.05$ and $f_{\rm ov} = 0.035$ in Fig.~\ref{fig:nu_rot_M130}, come from variations in the surface rotation. These variations are of numerical origin, and increasing the temporal or spatial resolution of the model does not always solve this. Therefore, only the global trends can be trusted. 
It should also be noted that the CBM efficiency (and thus the amount of time the star spends on the MS) greatly influences the level of differentiality when AM transport is relatively weak. More efficient CBM results in a larger ratio of $\Omega_{\rm gc}/\Omega_{\rm surf}$ at the same \xc, compared to when CBM is less efficient, as can been seen in Figs.~\ref{fig:nu_rot_M130} and \ref{fig:nu_rot_M200} for example. The models with rotationally-induced AM transport predict a weaker shear the envelope compared to the models with a uniform viscosity, which explains why the 1.3\dMsun models with a large CBM region as shown in Fig.~\ref{fig:nu_rot_M130} (dashed-dotted lines) evolve differently compared to those shown in Fig.~\ref{fig:nu_nonrot_M130_O10}.  

\section{Rotation rate at the ZAMS} \label{sec:Omegai}
\begin{figure*}
    \centering
    \includegraphics[width = \textwidth]{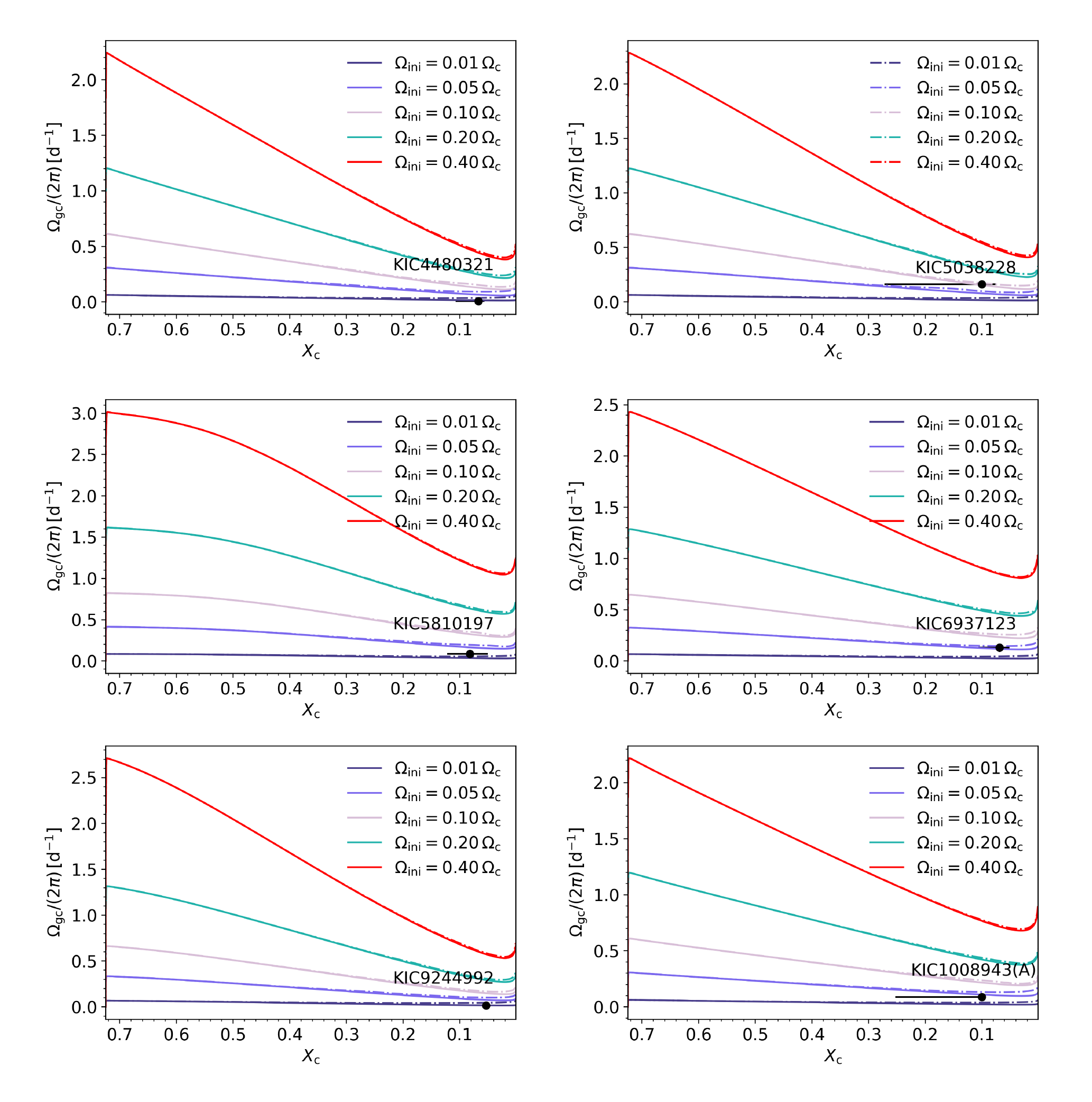}
    \caption{The predicted evolution of the g-mode cavity rotation frequency as a function of the hydrogen-mass fraction in the core for different initial rotation frequencies as a fraction of the initial Keplerian critical break-up frequency. In each panel, the stellar mass and \fov corresponds to those in the best-matching model listed in Table~\ref{tab:theta_sample}. The black markers indicate the observations, where the \xc estimate is from this work, and the rotation frequency from \citep[][uncertainties on $\Omega_{\rm gc}$ smaller than symbol size]{Li2019}. The solid lines are models with a constant uniform viscosity (Sect.~\ref{sec:nu_nonrot}), whereas the dashed-dotted lines are the models with rotational instabilities (Sect.~\ref{sec:nu_rot}). }
    \label{fig:Oi}
\end{figure*}

Using the stellar mass, $M_\star$, hydrogen-mass fraction, \xc, and CBM parameter, \fov, from the best-matching models of each star, we explore what the initial rotation rate near the ZAMS should be to reproduce the measured $\Omega_{\rm gc}$ from \cite{Li2019} at the current value of \xc. We do this exercise for the two treatments of AM transport discussed previously. In the case of a constant viscosity (i), its value is taken large enough to enforce uniform rotation along the MS. Thus, we set the viscosity to an arbitrary large value 
$\nu_{\rm AM,  non-rot} = 10^{20}~{\rm cm}^2\,{\rm s}^{-1}$. For each star, models with $(\Omega/\Omega_{\rm c})_{\rm i} \in [0.01, 0.05, 0.1, 0.2, 0.4]$ were computed. The goal here is to get a rough estimate of the initial conditions at birth. The solid lines in Fig.~\ref{fig:Oi} show the predicted evolution of the rotation frequency of the g-mode cavity of the best-matching model of each star for different initial rotation rates. The last column of Table~\ref{tab:theta_sample} lists the initial rotation frequency needed to best reproduce the current near-core rotation rate. When a large uniform viscosity is assumed that enforces uniform rotation, we find that the rotation frequency at birth for these six stars must be below 10~percent of the (initial) critical break-up frequency. Therefore, these slowly rotating stars must have been born with relatively little AM, or the transport of AM operating in these stars cannot be modelled by a uniform viscosity. As can be seen in Fig.~\ref{fig:Oi}, the largest rotation frequency at the ZAMS is about $0.6\,{\rm d}^{-1}$, or $7\,\mu{\rm Hz}$. This is consistent with the PMS models presented in \cite{Ouazzani2019} that are described by an accretion disk lifetime of 5\,Myr and a disc rotation period of 7.2\,d (see their Fig.~1). \newline 

\begin{figure}
    \centering
    \includegraphics[width = \columnwidth]{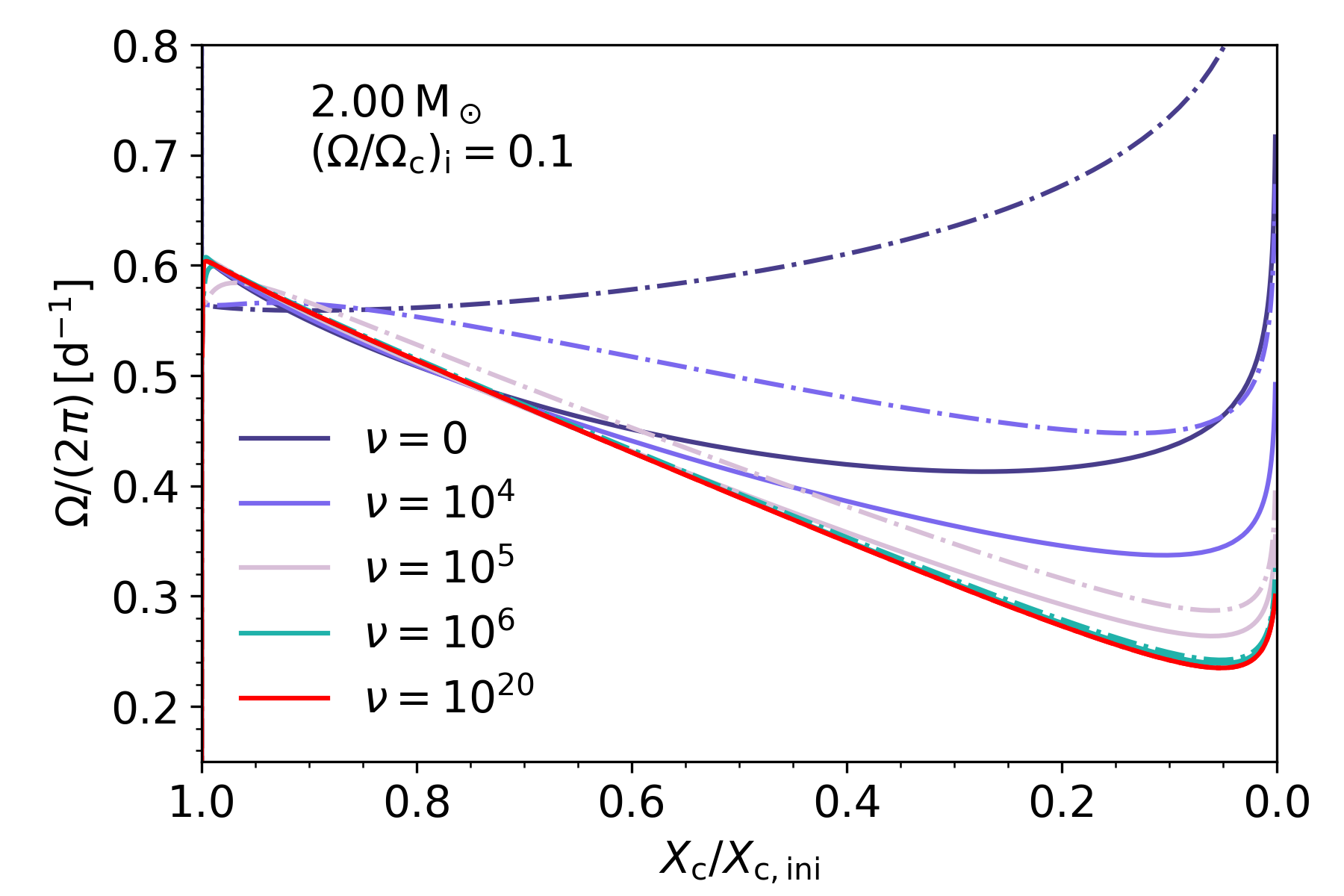}
    \caption{\edit{Evolution of the g-mode cavity rotation, $\Omega_{\rm gc}$ (solid lines), and the actual core rotation (dashed-dotted lines) of a $2.0\,{\rm M_\odot}$ model with $(\Omega/\Omega_{\rm c})_{\rm i} = 0.10$ and $f_{\rm ov} = 0.005$, computed for various values for the constant uniform viscosity (in ${\rm cm^2\,s^{-1}}$). For model with $\nu = 0$, there is no AM transport.}}
    \label{fig:visc_core}
\end{figure}

The same exercise is repeated for the models with AM transport induced by the processes listed in Sect.~\ref{sec:AM_trans}, without any additional scaling of the contributions. These models are shown as dashed-dotted lines in Fig.~\ref{fig:Oi}. While the level of radial differential rotation depends strongly on the initial rotation, which is in contrast to the models with a constant viscosity, the evolution of the rotation frequency of the g-mode cavity is similar. In Fig.~\ref{fig:visc_core}, the evolution of $\Omega_{\rm gc}$ for a 2\dMsun model is shown for different values for a uniform viscosity, including a model without any AM transport. This figure also shows the rotation frequency of the convective core.\footnote{It should be noted that $\Omega_{\rm gc}$, although often referred to as the core rotation rate, is actually a weighted rotation frequency just outside of the core.} For the inferred values of the uniform viscosity in this work ($>\,10^5\,{\rm cm^2\,s^{-1}}$), the value of $\Omega_{\rm gc}$ is close to the actual core rotation frequency. Moreover, when the viscosity is large enough such that the near-core region is corotating with the core, increasing it even more will not change the evolution of $\Omega_{\rm gc}$. The evolution of $\Omega_{\rm gc}$ predicted by the models shown in Fig.~\ref{fig:Oi} with a large uniform viscosity and the models with rotationally-induced processes is similar. This implies that the rotationally-induced processes are effective enough to enforce corotation of the near-core region (probed by $\Omega_{\rm gc}$) with the convective core.  As can been seen from the panels in Fig.~\ref{fig:Oi}, the estimation of the initial rotation frequency is similar to the one found by a constant viscosity. Yet, as discussed in Sect.~\ref{sec:nu_rot}, the very low rotation frequencies at the ZAMS inferred for KIC4480321, KIC5810197 and KIC9244992 (i.e. around one percent of the critical break-up frequency) results in core-to-surface rotation rates that are higher than what is observed (see Figs.~\ref{fig:nu_rot_M130} and \ref{fig:nu_rot_M200}). In case of KIC9244992, the core-to-surface rotation has been measured \citep{Saio2015}, while for \editf{the other two stars} we take the upper limit on the radial differential rotation of the sample studied by \cite{Saio2021}. It is therefore also possible that this upper limit is not representative for all F-type stars, and that KIC4480321 and KIC5810197 in reality actually have a much higher level of differential rotation than what has been observed in \gDor pulsators so far. We can thus conclude that AM transport via the rotationally-induced processes listed in Sect.~\ref{sec:AM_trans} can explain the observed rotation profiles of half of the stars in the sample, but is (potentially) overpredicting the level of radial differential rotation in some stars that seem to have been born with a very low rotation rate. To reconcile this with these observations, the factor $f_{\rm \nu, rot}$ would need to be increased to about a factor 1000 to ensure $\Omega_{\rm gc}/\Omega_{\rm surf} \lesssim 1.1$ on the MS, as shown the Fig.~\ref{fig:nu_rot_f1000} in the Appendix.  

\section{Conclusion} \label{sec:conclusions}
This paper presents a test of transport processes of AM in intermediate-mass stars, using the open-source stellar evolution code \mesa. First, forward asteroseismic was performed on seven slowly rotating \gDor pulsators from \cite{Li2019}, following the method by \cite{Mombarg2021}. The stellar mass, age, and core-boundary mixing efficiency (or equivalently, the core mass) were derived from the observed g-mode periods with know mode identification. Next, two treatments of AM transport were considered; (i) a constant uniform viscosity, and (ii) rotationally-induced processes (including the Spruit-Tayler dynamo).
In case (i), the aim was to calibrate what order of magnitude the viscosity needs to be, while in case (ii) the aim was to conclude whether the physical implementation of the processes thought to be responsible for the transport of AM is indeed adequate. This approach differs slightly from other studies testing AM transport in stars, where an ad-hoc constant uniform viscosity is added to the one predicted from physical processes to account for any missing physics \citep[e.g.][]{Eggenberger2012, Ouazzani2019}. In the stellar evolution models with AM transport, the physics regarding the rotation on the pre-main sequence are included in a free parameter that is the uniform rotation frequency at the start of the main sequence. This initial rotation frequency was inferred for six stars, using the present-day near-core rotation frequency and the asteroseismically derived mass and age. The inferred initial rotation is only weakly dependent on the assumed treatment of AM transport, and for both for prescriptions (i) and (ii), initial rotation rates below 10 percent of the initial critical rotation frequency were found. For treatment (i), we find the uniform viscosity to be somewhere between roughly $2\cdot 10^5~{\rm cm}^2\,{\rm s}^{-1}$ and $5\cdot 10^7~{\rm cm}^2\,{\rm s}^{-1}$ such that it reproduces the observed core-to-surface rotation ratios, dependent on the stellar mass and core-boundary mixing efficiency. Models with only rotationally-induced processes (ii) can reproduce the observed rotation profiles in \gDor pulsators if the initial rotation rates at the ZAMS are higher than about five percent of the initial critical break-up frequency. However, for the initial rotation frequency inferred for three out of the six stars in this work, KIC4480321, KIC5810197, and KIC9244992, the ratio of the core-to-surface rotation frequency near the end of the main sequence (where the stars are situated) is predicted to be larger than what has been observed. For KIC9244992 and KIC10080943(A) this ratio has been inferred by previous works, while for the other four stars we take results from other samples as an upper limit \citep[e.g.][]{VanReeth2018, Li2020, Saio2021, Schmid2016}. Therefore, it is possible that KIC4480321 and KIC5810197 actually have a larger core-to-surface ratio than what has been observed in \gDor pulsators so far. However, assuming that these two stars are no outliers, is seems that AM transport by rotationally-induced processes, while adequate for three out of six stars, is too inefficient for a significant part of the sample. 

\editt{\cite{Ouazzani2019} performed a sample study of \gDor stars and their near-core rotation frequencies, using the buoyancy travel time as an age indicator.} The physics concerning the AM transport used in this work are similar to that of theirs, except that here, a diffusive approach of Eddington-Sweet circulation is used, and the Spruit-Tayler dynamo is included. While \cite{Ouazzani2019} conclude AM transport by meridional circulation and shear-induced turbulence following the work of \cite{Zahn1992} not adequate, the physics tested in this paper can explain the core and surface rotation frequencies of four out of the six stars. This conclusion is supported by the recent findings that AM transport models with magnetism are favoured in intermediate-mass stars, compared to those with only hydrodynamical processes \citep[][\editt{excluding internal gravity waves}]{Moyano2023}. Here, we also find that the Spruit-Tayler dynamo is a dominant process in the transport of AM from the processes listed in Sect.~\ref{sec:AM_trans}. The inclusion of AM transport by internal gravity waves \citep[a non-diffusive process,][]{Mathis2013} could be a possible way to resolve any discrepancies between the observations and models in some stars. 

The stars presented in this paper (except KIC6937123) have been selected based on their high potential for rotation inversion thanks to their mode splittings, providing even more stringent constraints on the AM transport than only the rotation of the g-mode cavity and the surface. On the theoretical side, 2D stellar models that account for the centrifugal deformation and deliver the rotation profile in a self-consistent manner \citep{EspinosaLara2013, Rieutord2016} will help us better understand the interaction of the convective core with the radiative envelope.

\begin{acknowledgements}
  The author thanks Tami Rogers, Seth Gossage and S\'ebastien Deheuvels for the discussions, and Conny Aerts, Ashlin Varghese and Michel Rieutord for their comments on the manuscript. The author is also grateful for the helpful comments of the anonymous referee.
  The research leading to these results has received funding from the KU\,Leuven Research Council (grant C16/18/005: PARADISE), and from the French Agence Nationale de la Recherche (ANR), under grant MASSIF (ANR-21-CE31-0018-02). The computational resources and services used in this work were provided by the VSC (Flemish Supercomputer Center), funded by the Research Foundation - Flanders (FWO) and the Flemish Government department EWI. This research made use of the \texttt{numpy} \citep{Harris2020} and \texttt{matplotlib} \citep{Hunter2007} \texttt{Python} software packages.   
\end{acknowledgements}

\bibliographystyle{aa} % style aa.bst
\bibliography{main} % your references Yourfile.bib

\appendix

\section{Best-fitting parameters from neural network}
\begin{table}[h!]
    \centering
    \caption{Best-fitting parameters predicted by the \ctpo neural network.}
    \begin{tabular}{llll}
    \hline \hline
    KIC & $M_\star\,[\rm M_\odot]$ & $X_{\rm c}$ & $f_{\rm ov}$ \\
    \hline
    2450944 & $1.309_{-0.009}^{+0.690}$ & $0.638_{-0.585}^{+0.062}$ & $0.0282_{-0.0182}^{+0.0018}$ \\

3127996 & $1.651_{-0.183}^{+0.047}$ & $0.198_{-0.023}^{+0.227}$ & $0.0219_{-0.0119}^{+0.0081}$ \\
\textbf{4480321} & $1.997_{-0.027}^{+0.003}$ & $0.067_{-0.011}^{+0.040}$ & $0.0299_{-0.0086}^{+0.0001}$ \\
4919344 & $1.511_{-0.044}^{+0.174}$ & $0.426_{-0.253}^{+0.214}$ & $0.0293_{-0.0193}^{+0.0007}$ \\
\textbf{5038228} & $1.853_{-0.132}^{+0.093}$ & $0.186_{-0.111}^{+0.086}$ & $0.0101_{-0.0001}^{+0.0199}$ \\
5459805 & $1.953_{-0.159}^{+0.047}$ & $0.187_{-0.037}^{+0.262}$ & $0.0298_{-0.0198}^{+0.0002}$ \\
5557072 & $1.308_{-0.008}^{+0.436}$ & $0.314_{-0.264}^{+0.100}$ & $0.0117_{-0.0017}^{+0.0161}$ \\
\textbf{5810197} & $1.308_{-0.007}^{+0.203}$ & $0.082_{-0.032}^{+0.040}$ & $0.0297_{-0.0197}^{+0.0002}$ \\
6302589 & $1.737_{-0.145}^{+0.080}$ & $0.236_{-0.008}^{+0.246}$ & $0.0295_{-0.0195}^{+0.0004}$ \\
6467639 & $1.651_{-0.029}^{+0.339}$ & $0.182_{-0.132}^{+0.045}$ & $0.0220_{-0.0118}^{+0.0080}$ \\
\textbf{6937123} & $1.570_{-0.006}^{+0.155}$ & $0.069_{-0.018}^{+0.037}$ & $0.0243_{-0.0141}^{+0.0029}$ \\
7661054 & $1.475_{-0.097}^{+0.107}$ & $0.533_{-0.189}^{+0.161}$ & $0.0197_{-0.0096}^{+0.0093}$ \\
7697861 & $1.603_{-0.035}^{+0.145}$ & $0.525_{-0.191}^{+0.138}$ & $0.0300_{-0.0180}^{+0.0000}$ \\
9028134 & $1.566_{-0.076}^{+0.166}$ & $0.083_{-0.032}^{+0.226}$ & $0.0100_{-0.0000}^{+0.0199}$ \\
\textbf{9244992} & $1.637_{-0.020}^{+0.000}$ & $0.053_{-0.003}^{+0.004}$ & $0.0101_{-0.0000}^{+0.0025}$ \\
\textbf{10080943(A)} & $1.851_{-0.051}^{+0.149}$ & $0.177_{-0.082}^{+0.076}$ & $0.0275_{-0.0129}^{+0.0025}$ \\
\textbf{10080943(B)} & $1.953_{-0.153}^{+0.047}$ & $0.127_{-0.077}^{+0.086}$ & $0.0105_{-0.0005}^{+0.0186}$ \\
10468883 & $1.529_{-0.101}^{+0.147}$ & $0.373_{-0.205}^{+0.245}$ & $0.0102_{-0.0002}^{+0.0198}$ \\

\hline
    \end{tabular}
    \tablefoot{The stars that are modelled in more detail in this paper are indicated in bold face. }
    \label{tab:theta_NN}
\end{table}

\section{Period-spacing patterns }
In this appendix, the predicted period-spacing patterns corresponding to the best matching models listed in Table~\ref{tab:theta_sample} are shown.
The uncertainties on the observations are smaller than the symbol size. 

\begin{figure}
    \centering
    \includegraphics[width = \columnwidth]{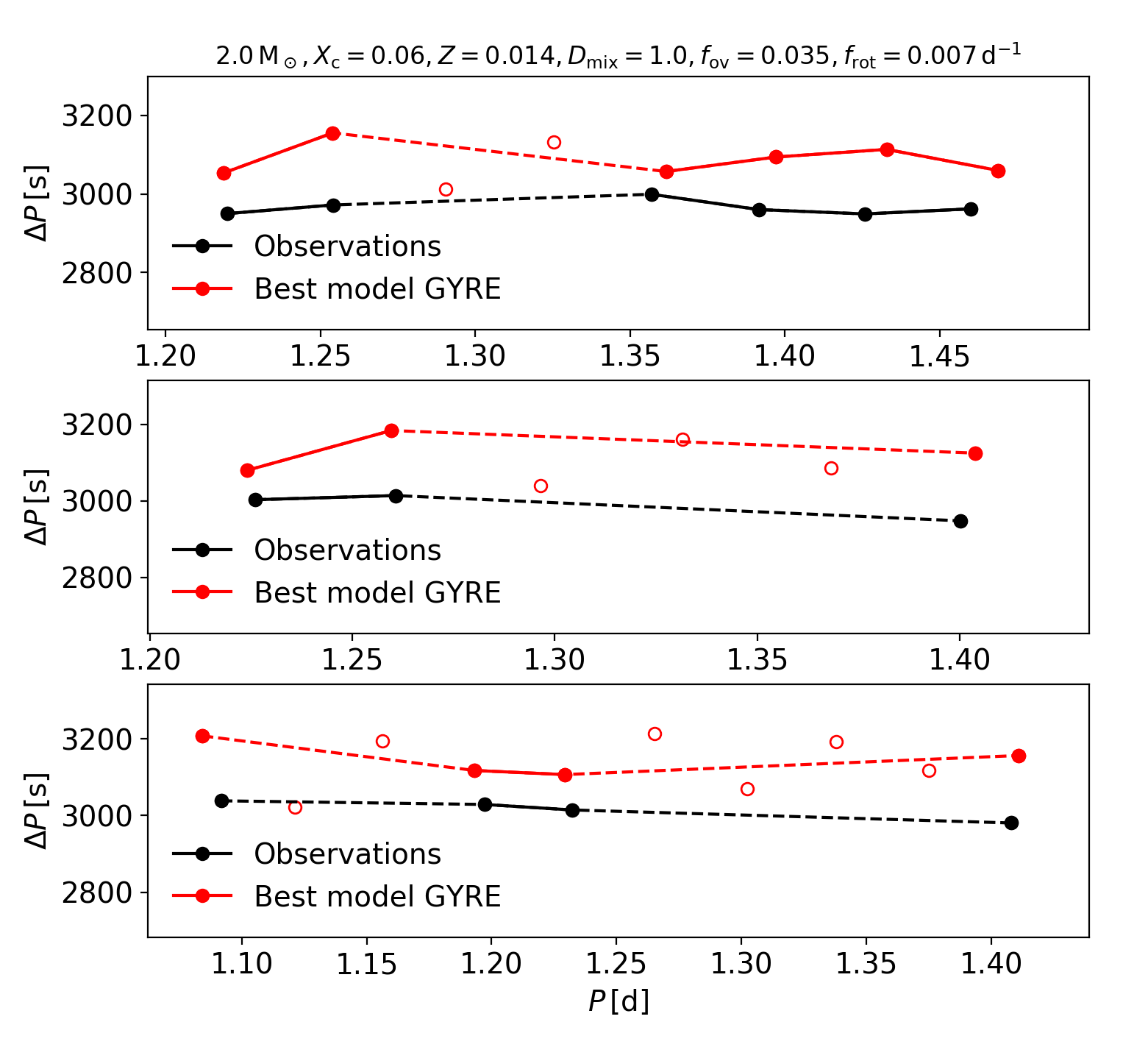}
    \caption{Period-spacing patterns of the best-matching \gyre model for KIC4480321 (in red) and the observed pattern from \cite{Li2019} (in black). Open symbols indicate skipped radial orders in the model. Mode IDs $(\ell, m)$ from top to bottom: (1,1), (1,0), and $(1,-1)$. }
    \label{fig:PSP44}
\end{figure}

\begin{figure}
    \centering
    \includegraphics[width = \columnwidth]{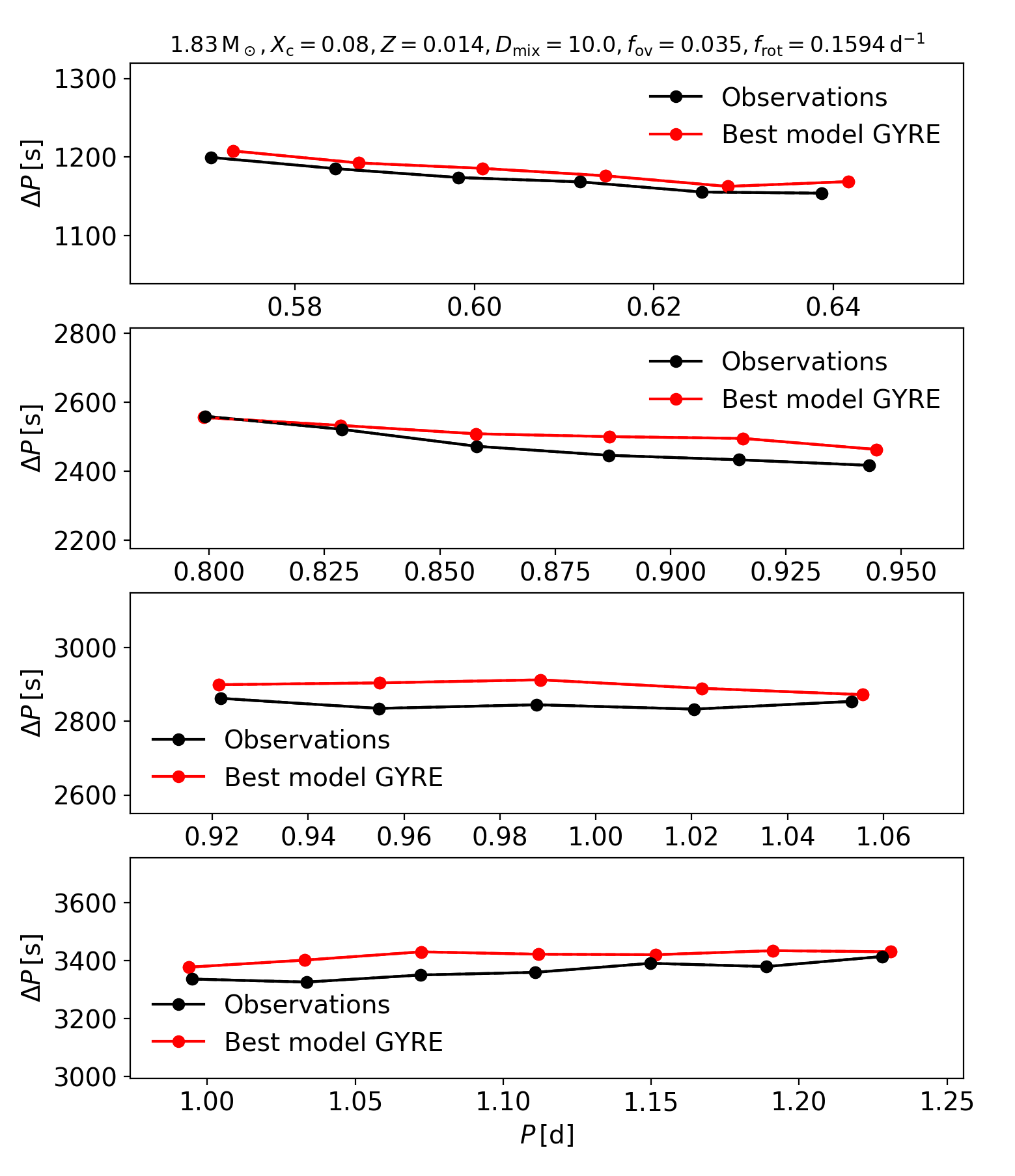}
    \caption{Period-spacing patterns of the best-matching \gyre model for KIC5038228 (in red) and the observed pattern from \cite{Li2019} (in black). Mode IDs $(\ell, m)$ from top to bottom: (2,2), (1,1), (1,0), and $(1,-1)$.}
    \label{fig:PSP50}
\end{figure}

\begin{figure}
    \centering
    \includegraphics[width = \columnwidth]{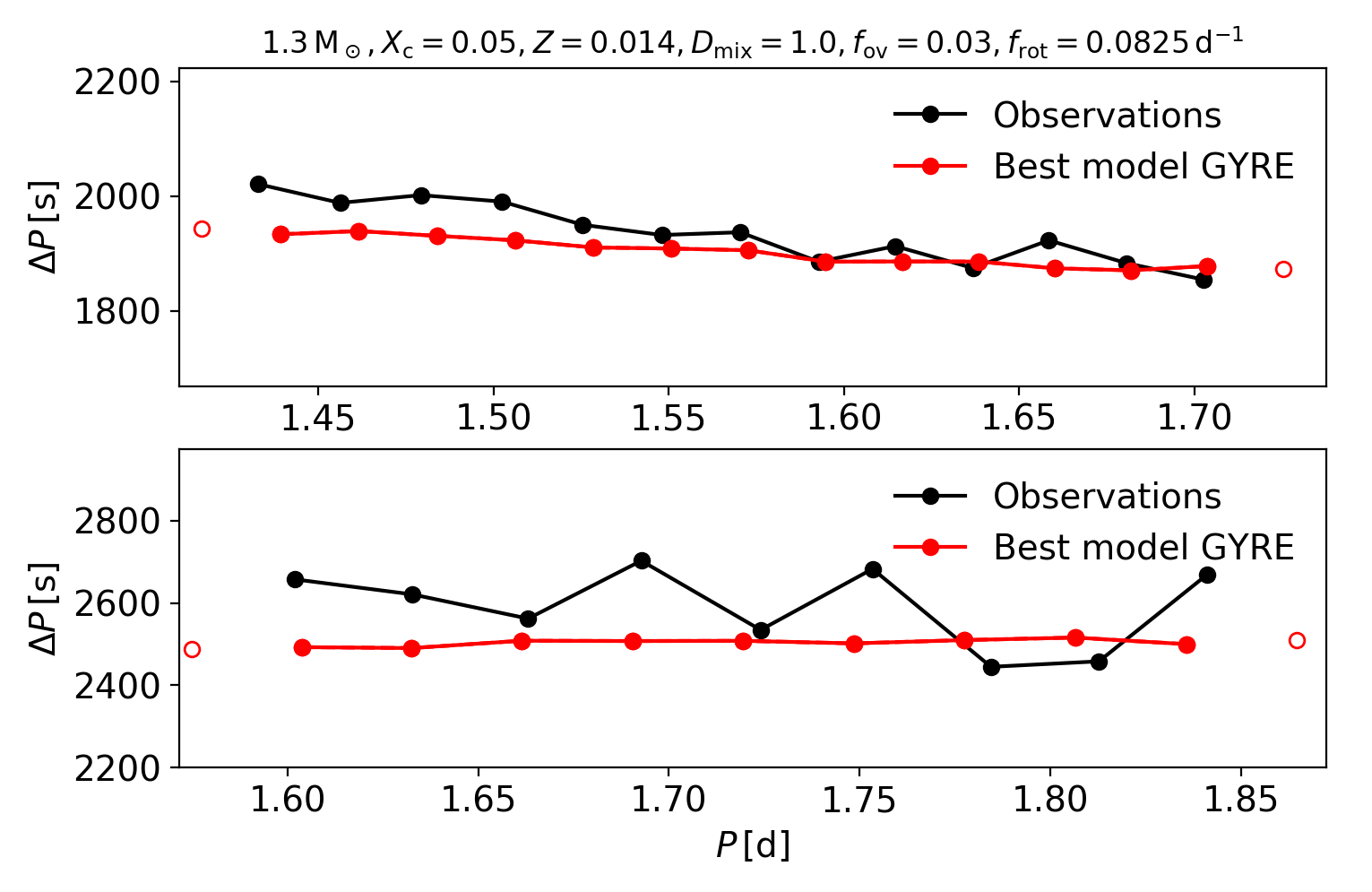}
    \caption{Period-spacing patterns of the best-matching \gyre model for KIC5810197 (in red) and the observed pattern from \cite{Li2019} (in black). Open symbols indicate skipped radial orders in the model. Mode IDs $(\ell, m)$ from top to bottom: (1,1) and $(1,-1)$.}
    \label{fig:PSP58}
\end{figure}

\begin{figure}
    \centering
    \includegraphics[width = \columnwidth]{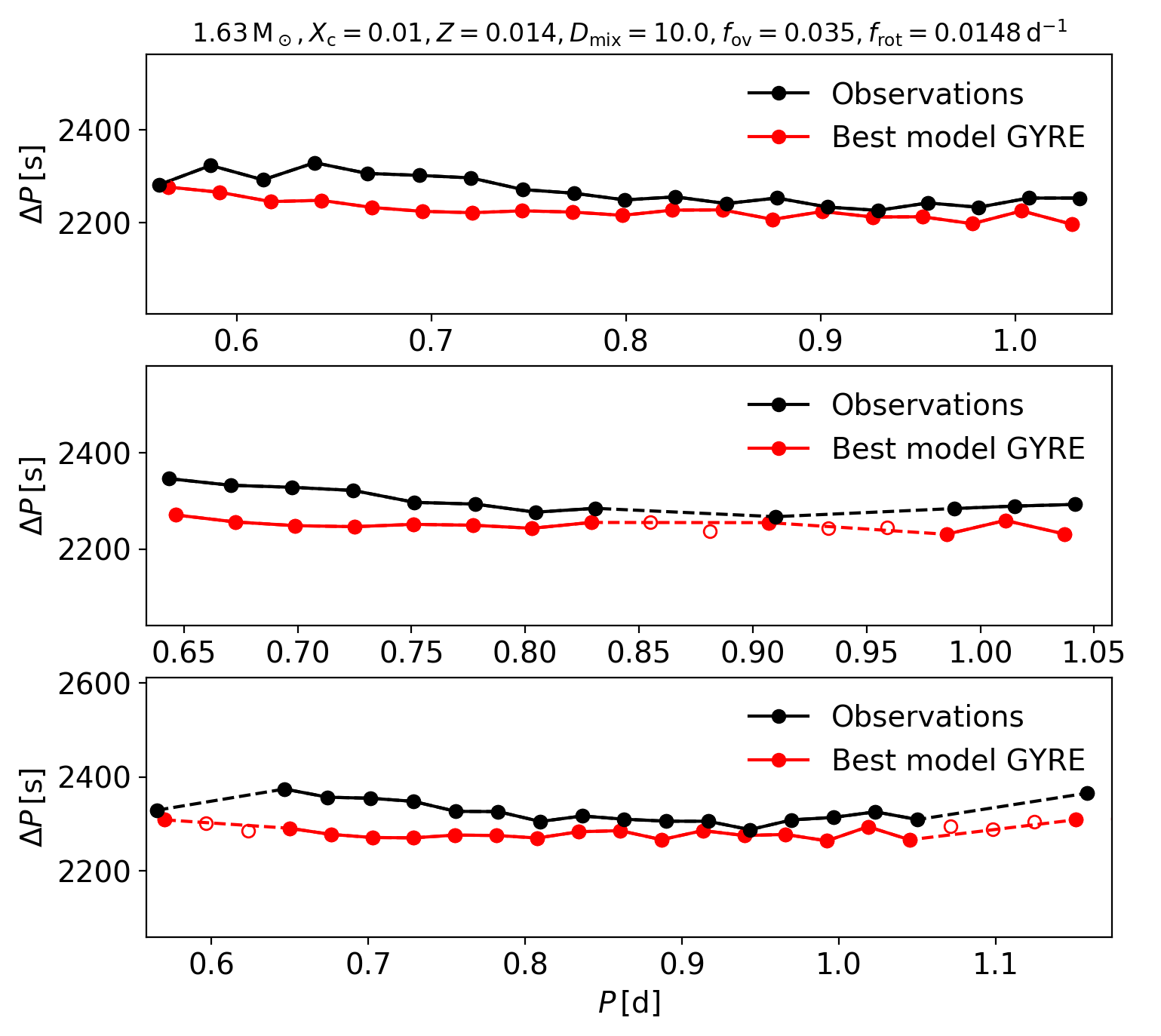}
    \caption{Period-spacing patterns of the best-matching \gyre model for KIC9244992 (in red) and the observed pattern from \cite{Li2019} (in black). Open symbols indicate skipped radial orders in the model. Mode IDs $(\ell, m)$ from top to bottom: (1,1), (1,0), and $(1,-1)$.}
    \label{fig:PSP92}
\end{figure}

\begin{figure}
    \centering
    \includegraphics[width = \columnwidth]{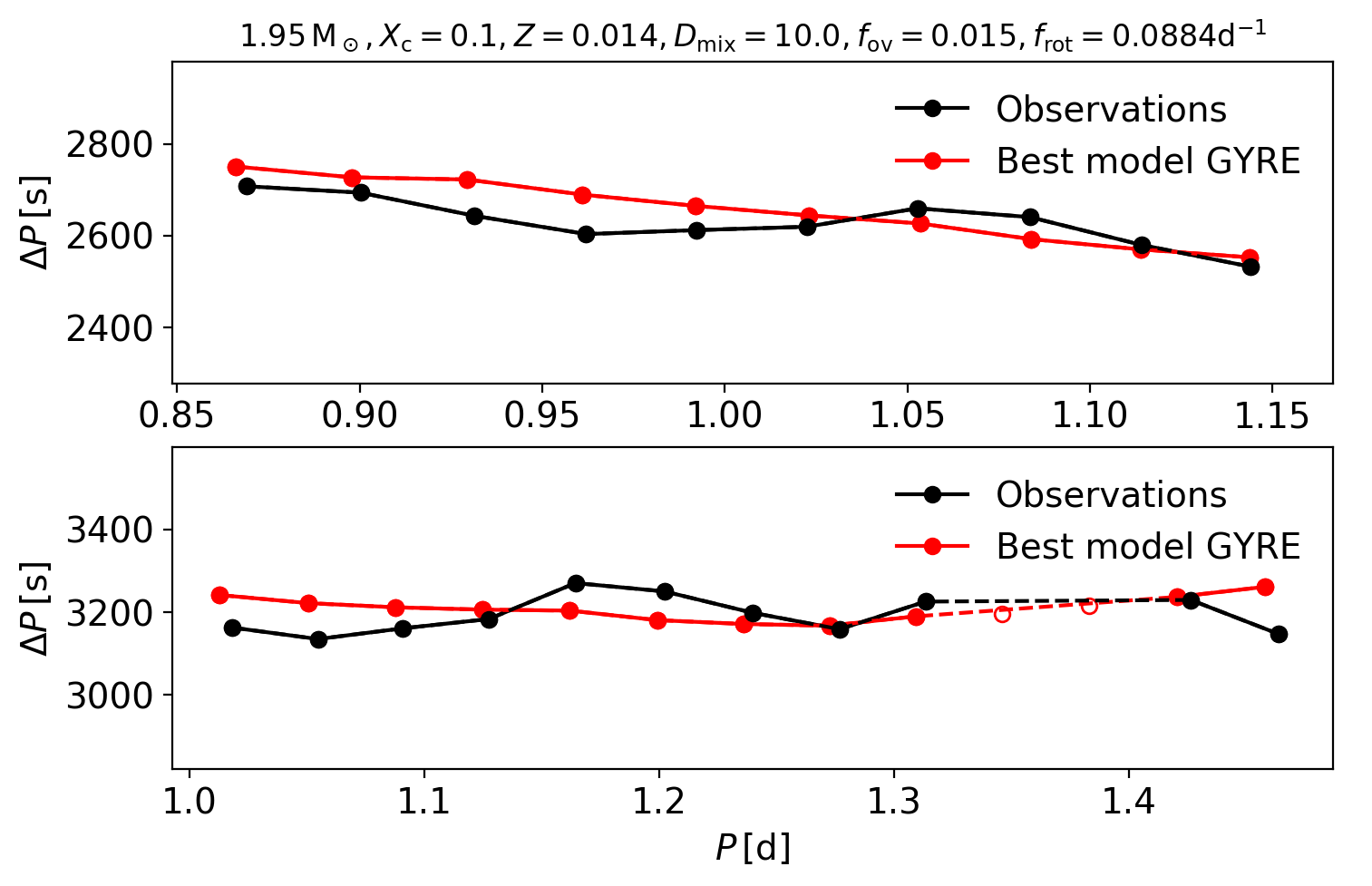}
    \caption{Period-spacing patterns of the best-matching \gyre model for KIC10080943(A) (in red) and the observed pattern from \cite{Li2019} (in black). Open symbols indicate skipped radial orders in the model. Mode IDs $(\ell, m)$ from top to bottom: (1,1), and $(1,-1)$.}
    \label{fig:PSP10A}
\end{figure}

\begin{figure}
    \centering
    \includegraphics[width = \columnwidth]{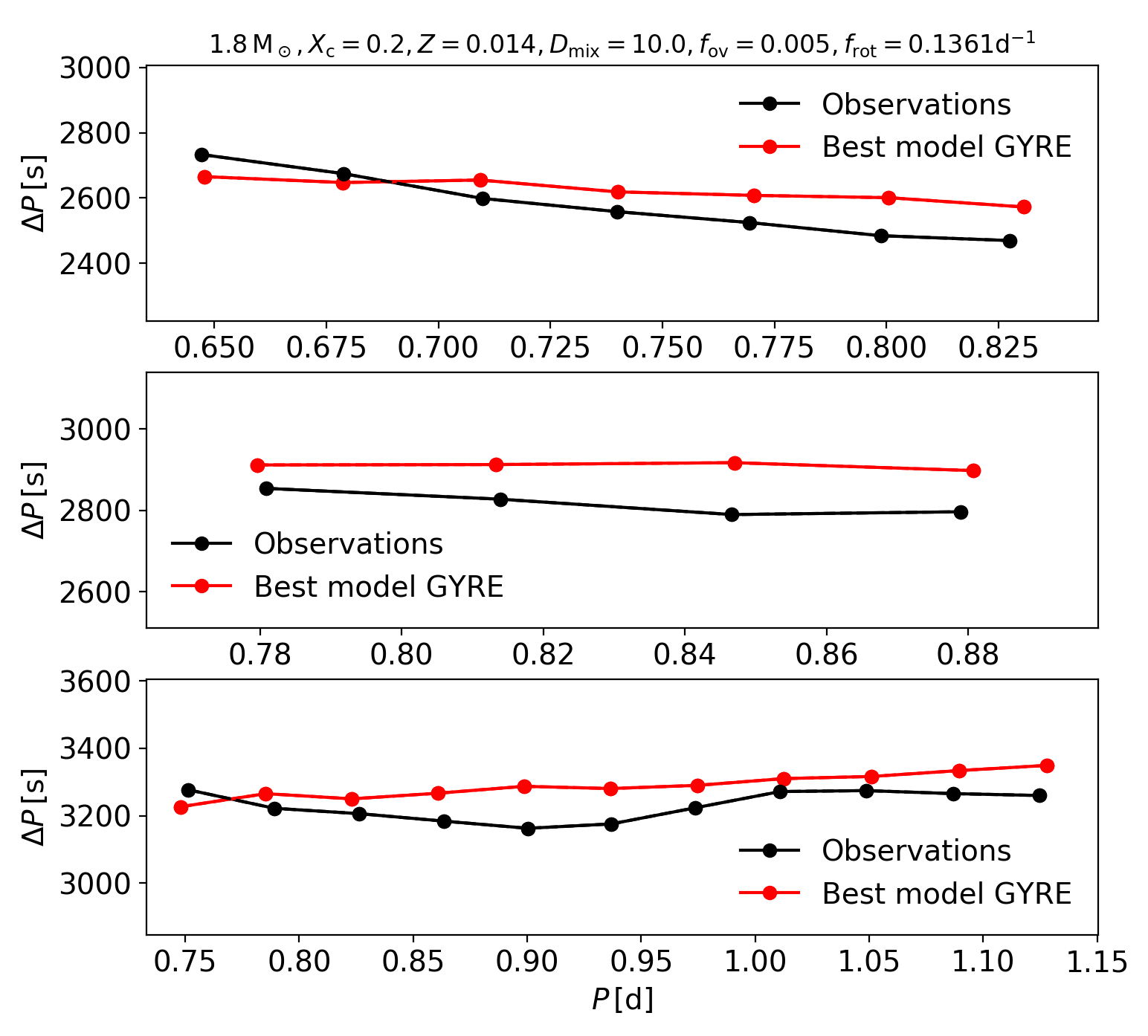}
    \caption{Period-spacing patterns of the best-matching \gyre model for KIC10080943(B) (in red) and the observed pattern from \cite{Li2019} (in black). Mode IDs $(\ell, m)$ from top to bottom: (1,1), (1,0), and $(1,-1)$.}
    \label{fig:PSP10B}
\end{figure}

\section{Influence of the initial rotation frequency}
In this appendix, the evolution of the core-to-surface rotation frequency ratio is shown, similar to Figs.~\ref{fig:nu_nonrot_M130_O10} and \ref{fig:nu_nonrot_M200_O10}, but for a higher initial rotation $(\Omega/\Omega_{\rm c})_{\rm i} = 0.4$.
\begin{figure}
    \centering
    \includegraphics[width = \columnwidth]{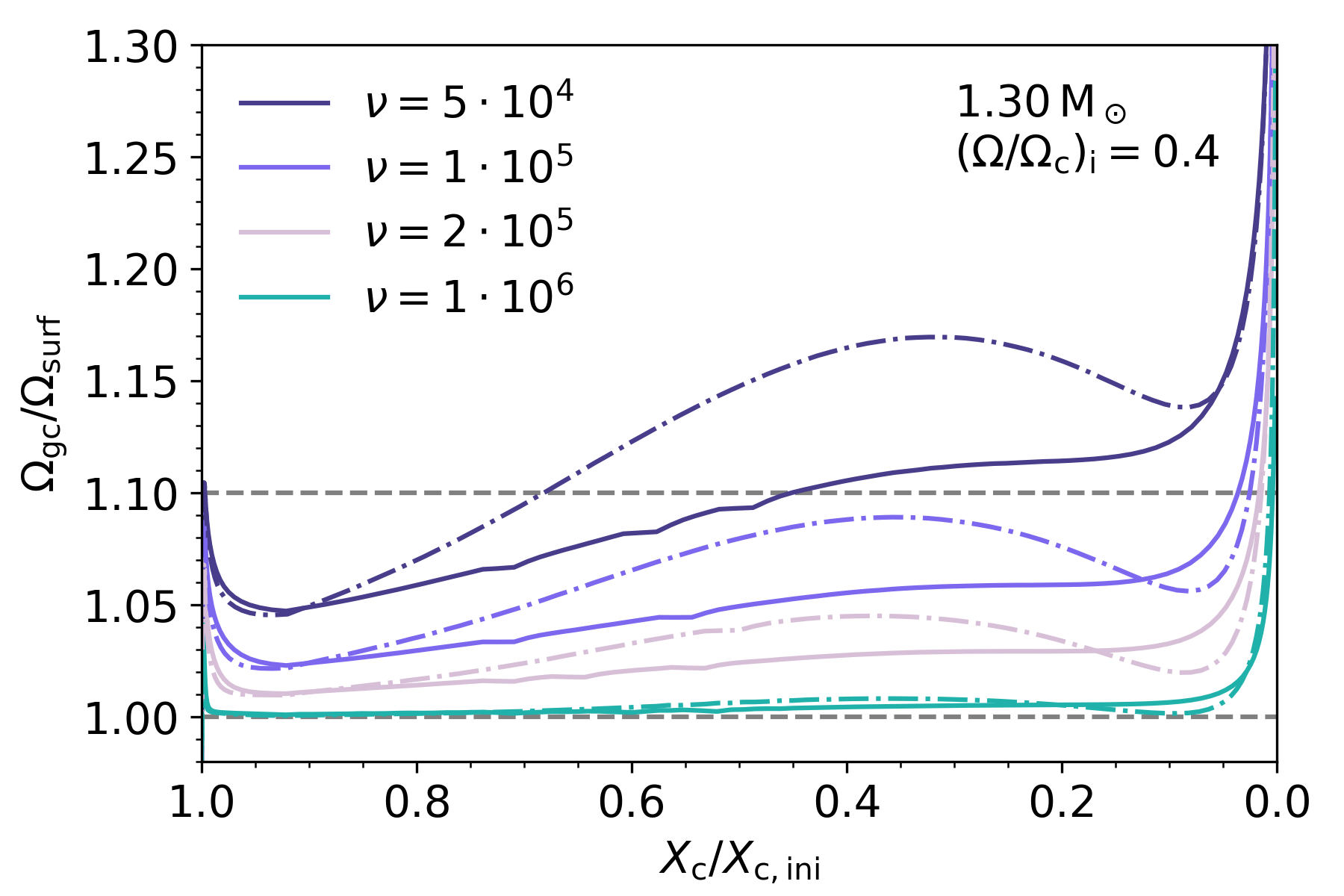}
    \caption{The predicted ratio of the g-mode cavity rotation frequency over the surface rotation frequency as a function of evolution for a 1.3\dMsun model starting with $(\Omega/\Omega_{\rm c})_{\rm i} = 0.4$. In these models, AM transport is modelled by a constant uniform viscosity, shown in the legend (in cm$^{2}$\,s$^{-1}$). The solid lines are for $f_{\rm ov} = 0.005$, the dashed-dotted lines for $f_{\rm ov} = 0.035$. The typical upper limit of the observed range of core-to-surface rotation is shown by a grey dashed line.}
    \label{fig:nu_nonrot_M130_O40}
\end{figure}

\begin{figure}
    \centering
    \includegraphics[width = \columnwidth]{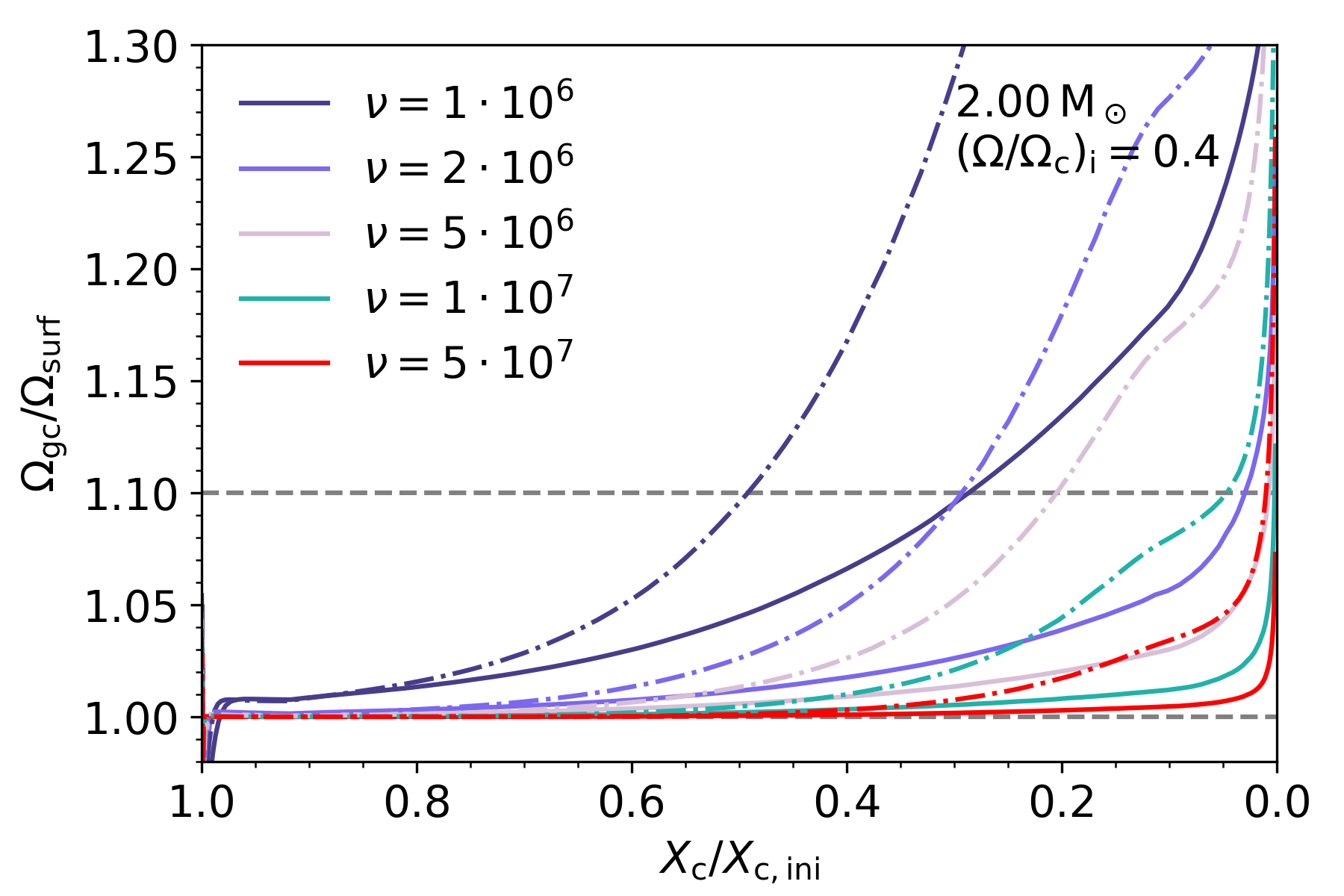}
    \caption{Same as Fig.~\ref{fig:nu_nonrot_M130_O40}, but for a 2\dMsun model.}
    \label{fig:nu_nonrot_M200_O40}
\end{figure}

\section{Boosted viscosity for rotationally-induced processes}
In this appendix, the predicted core-to-surface rotation frequency ratio is shown when $f_{\rm \nu, rot} = 10^3$ (See Eq.~(\ref{eq:nu_rot})). This higher factor is needed to reconcile the predicted rotation profiles with the observed ones, if it is assumed the rotationally-induced processes listed in Sect.~\ref{sec:AM_trans} are the only source of AM transport. 

\begin{figure}
    \centering
    \includegraphics[width = \columnwidth]{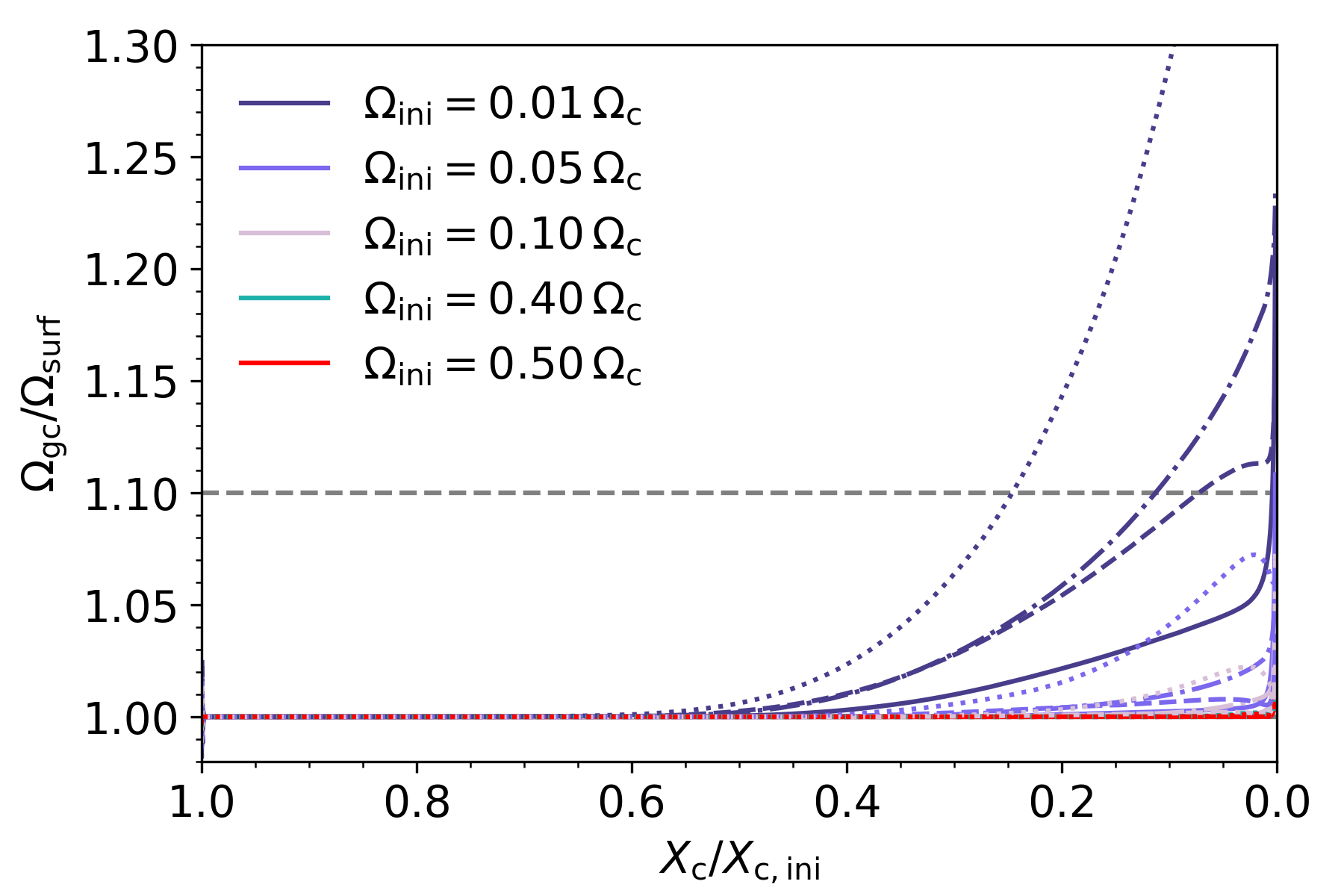}
    \caption{The predicted ratio of the g-mode cavity rotation frequency over the surface rotation frequency as a function of evolution for $f_{\rm \nu, rot} = 1000$. In these models, AM transport is computed from the rotationally-induced processes described in Eq~(\ref{eq:nu_rot}). The solid lines are for ($1.3\,{\rm M_\odot},~f_{\rm ov} = 0.005$), the dashed-dotted lines for ($1.3\,{\rm M_\odot},~f_{\rm ov} = 0.035$), the dashed lines for ($2.0\,{\rm M_\odot},~f_{\rm ov} = 0.005$), and the dotted lines for ($2.0\,{\rm M_\odot},~f_{\rm ov} = 0.035$). The typical upper limit of the observed range of core-to-surface rotation is shown by the upper horizontal grey dashed line. The lower grey dashed line indicates solid body rotation.}
    \label{fig:nu_rot_f1000}
\end{figure}

\end{document}